\documentclass[iop]{emulateapj} 
\usepackage{color}
\usepackage{amsmath}
\def\deg{^{\circ}}
\newcommand{\bdma}[1]{\mbox{\boldmath $#1$}}

\newcommand{\be}{\begin{eqnarray}}
\newcommand{\ee}{\end{eqnarray}}
\newcommand{\ra}{{\bf r_A}}
\newcommand{\rb}{{\bf r_B}}
\newcommand{\ybz}{{\rm y_B(0)}}
\newcommand{\va}{{\bf V_A}}
\newcommand{\vb}{{\bf V_B}}

\newcommand{\ta}{t_A}
\newcommand{\tb}{t_B}

\newcommand{\siss}{s_0}
\newcommand{\sisp}{s_p}
\newcommand{\tiss}{T_{\rm ISS}}

\newcommand{\vcx}{V_{\rm Cx}}
\newcommand{\vcy}{V_{\rm Cy}}

\newcommand{\vpa}{V_{\rm P\alpha}}
\newcommand{\vpd}{V_{\rm P\delta}}

\newcommand{\ar}{A_R}

\newcommand{\Ko}{K_0}
\newcommand{\Ks}{K_S}
\newcommand{\Kss}{K_{S2}}
\newcommand{\Kc}{K_{C}}
\newcommand{\Kcc}{K_{C2}}

\newcommand{\vsigma}{{ \mbox{\boldmath $\sigma$} }}

\def\simless{\mathbin{\lower 3pt\hbox
   {$\rlap{\raise 5pt\hbox{$\char'074$}}\mathchar"7218$}}} 
\def\simgreat{\mathbin{\lower 3pt\hbox
   {$\rlap{\raise 5pt\hbox{$\char'076$}}\mathchar"7218$}}} 

\begin{document}

\title{Interstellar Scintillation of the Double Pulsar J0737$-$3039}
\author{B. J. Rickett\altaffilmark{1}, 
W. A. Coles\altaffilmark{1}, 
C. F. Nava\altaffilmark{1},
M. A. McLaughlin\altaffilmark{2,3},  
S. M. Ransom\altaffilmark{4}, \\
F. Camilo\altaffilmark{7,8},
R. D. Ferdman\altaffilmark{3,6}, 
P. C. C. Freire\altaffilmark{6,9}, 
M. Kramer\altaffilmark{3,9},
A. G.  Lyne\altaffilmark{3},
I. H. Stairs\altaffilmark{5}
}

\altaffiltext{1}{ECE Dept., University of California San Diego,  La Jolla, CA 92093-0407, USA; bjrickett@ucsd.edu}
\altaffiltext{2}{West Virginia University, Morgantown, WV 26505, USA}
\altaffiltext{3}{Jodrell Bank Center for Astrophysics, University of Manchester, M13 9PL, UK} 
\altaffiltext{4}{National Radio Astronomy Observatory, Charlottesville, VA 22903, USA}
\altaffiltext{5}{Dept. of Physics and Astronomy, University of British Columbia, Vancouver, BC V6T 1Z1, Canada}
\altaffiltext{6}{Dept. of Physics, McGill University, Montr\'{e}al, QC H3A 2T8, Canada}
\altaffiltext{7}{National Astronomy and Ionosphere Center, Arecibo, PR 00612-8346}
\altaffiltext{8}{Columbia Astrophysics Laboratory, Columbia University, New York, NY 10027}
\altaffiltext{9}{Max Planck Institut f\"{u}r Radioastronomie, Auf dem H\"{u}gel 69, D-53121 Bonn, Germany}

\begin{abstract}

We report here a series of observations of the interstellar
scintillation (ISS) of the double pulsar J0737$-$3039 over the
course of 18 months.  As in earlier work \citep{col05} the basic phenomenon
is the variation in the ISS caused by the changing transverse velocities
of each pulsar, the ionized interstellar medium (IISM), and the Earth. The transverse velocity
of the binary system can be determined both by VLBI and timing observations. The orbital 
velocity and inclination is almost completely determined from timing observations, but the direction of the 
orbital angular momentum is not known. Since the Earth's velocity is known, and can be 
compared with the orbital velocity by its effect on the timescale of the ISS, we can determine 
the orientation $\Omega$ of the pulsar orbit with respect to equatorial coordinates
($\Omega = 65\pm2\deg$). 
We also resolve the ambiguity ($i= 88.7$ or $91.3\deg$) in the inclination of the orbit deduced 
from the measured Shapiro delay by our estimate $i=88.1\pm0.5\deg$.
This relies on analysis of the ISS over both frequency and time and provides a
model for the location, anisotropy, turbulence level and transverse phase gradient of the IISM.
We find that the IISM can be well-modeled during each observation, typically of
a few orbital periods, but its turbulence level and mean velocity
vary significantly over the 18 months.

\end{abstract}


\clearpage

\section{Introduction}
The double pulsar binary system J0737$-$3039 is in
a highly relativistic orbit with significant eccentricity \citep{lyne04}. 
It is an eclipsing binary that is a wonderful laboratory for
studies of general relativity \citep{kramer06}. 
Detailed measurements of the eclipses of A have been used to probe the
magnetosphere of the B neutron star \citep{mcl04b, lyutikov}
and provide a measurement of geodetic precession \citep{breton08}. 
The changes in the pulse profiles 
have been used to explore the precession
of the emission beams, the evolution of the orbital system  
and the dynamics of B's supernova \citep{stairs06, ferdman, per10, per12}.  It is also a fine system for
study of the interstellar plasma (IISM) because the scattering is dominated
by a compact region, the velocities are well determined, and the plasma
turbulence can be probed on two neighbouring lines of sight.

\citet{ransom04} first reported how the interstellar intensity scintillation (ISS)
of the A pulsar exhibits dramatic modulation in timescale over its
orbital period (2.45 hr).  Following the method proposed by \citet{lyne84}
and developed by \citet{ord02} 
for PSR J1141$-$6545, the authors estimated a rather high center of mass 
velocity for the double pulsar.  
Subsequent analysis of the same data showed that
the scattering must be anisotropic and inclusion of this effect in the
analysis greatly reduced the implied center of mass velocity
\citep{col05} -- hereafter Paper 1. 
In this paper we found correlation between the ISS of pulsars 
A and B near the time of A's eclipse by B.  From the correlation we concluded
that the orbital inclination angle 
was considerably closer to 90$^\circ$ than had been expected on the 
basis of the original measurements of the Shapiro delay \citep{lyne04}.

The main purpose of the observations reported here was to make use of the
Earth's orbital velocity to improve, calibrate, and align the earlier
scintillation analyses. This would allow us to correct the center of mass
velocity for the motion of the Earth, to orient the binary orbit with
respect to the celestial reference frame, and to locate the distance
of the scattering region. The additional observations were also expected
to improve the estimates of the inclination of the orbit, the anisotropy
of the IISM and the spatial spectrum of the electron density of the IISM.
However, two factors made the original plan of analysis impossible. First
the phase at which emission from B is easily detectable drifted away from the 
time of A's eclipse during the course of the 2004-5 observations. This 
made measurements of the correlation between the A and B pulsars much less
consistent and reliable than had been expected. Second we observed that,
although the turbulence in the IISM is homogeneous over several binary 
orbits, it is not stationary over a year, nor is the velocity of the IISM constant
over the year. 
This phenomenon was also observed by Ord [private 
communication] when his group attempted to observe the effect of the
Earth's orbit on scintillations of PSR J1141$-$6545.

Subsequent pulse timing measurements have determined the 
Shapiro delay and proper motion with greater precision
\citep{kramer06}.  There have also been long baseline interferometry 
measurements of parallax and proper motion \citep{del09}.  
We have altered our analysis of the time scale
variations to take advantage of these observations, and we have modeled the
entire two dimensional time-frequency correlation function
of the scintillations, rather than simply
modeled its time scale. These changes have made it (just) 
possible to obtain a consistent interpretation. 
This provides the distance to the scattering region and the orientation of the 
pulsar orbit in celestial coordinates. It also provides an inclination estimate that 
is consistent with the Shapiro delay.

We now realize that the scattering is homogeneous over several binary orbits
because the proper motion of the pulsar is low and the binary orbit remains 
entirely within the ``scattering disc''. Since the measured intensity is a summation
over waves that have traveled through all possible paths through the scattering
disc, it is quite homogeneous over that area, even if the underlying turbulence is not.
However, from month to month, as we repeated the observations, different realizations
of the plasma turbulence occupied the scattering disc. The level of turbulence was 
clearly non-stationary on this time scale.  
  
\section{Observations and reduction to dynamic spectra}

Observations of the double pulsar system were made specifically for this
project with the 100-m Robert C. Byrd Green Bank Telescope (GBT) 
at intervals of 1-2 months from July 2004 to July 2005. 
All the observations were made with the SPIGOT auto-correlation spectrometer,
summing the polarizations every 81.92 $\mu$s \citep{kap05}.
On twelve of the epochs, observations were made with 1024 frequency channels over an 800 MHz 
bandwidth centered near 1900 MHz, however
only 600 MHz of the bandwidth was sufficiently free of radio frequency interference (RFI) 
to be useful. We have also been able to analyze data taken primarily for timing at 820 MHz 
with 1024 channels over a 50 MHz bandwidth on five epochs,
the first of which was analyzed in Paper 1. 
In total new and older observations span about 18 months
(see Table \ref{tab:harml} for dates).  We also analyzed 1400 MHz
dynamic spectra recorded earlier at Parkes \citep{manchester, burgay}.

The first step in the analysis of these data was to create
dynamic spectra for each pulsar in each frequency band. 
We edited the raw SPIGOT data files for RFI and Fourier transformed
the correlations with the Van Vleck corrections.
With SIGPROC we formed full pulse profiles at each frequency
with 64 and 256 phase bins, respectively, for pulsars A and B.
Individual frequency channels were
shifted with respect to each other using a dispersion measure of
48.9 cm$^{-3}$ pc \citep{lyne04}. 
These were added to create profiles at intervals of 10~s at 1900 MHz 
and 5~s or 10~s at 820 MHz. The pulse intensity from each profile was estimated
by first subtracting the average in an off-pulse window and then
integrating the profile to create a dynamic spectrum of pulse intensity versus
frequency and time.  
We estimated the gain in the passband using both the mean and rms in each
frequency channel. We found substantial variation in both measures over the
relatively broad bandwidths used. The rms appeared to be the better measure as
it was less affected by RFI, which was a serious problem at 1900 MHz.
Accordingly we corrected the gain of each channel by dividing it by the rms in
that channel.

\begin{figure*}
\includegraphics[trim=55 35 0 20,clip, width=18cm]{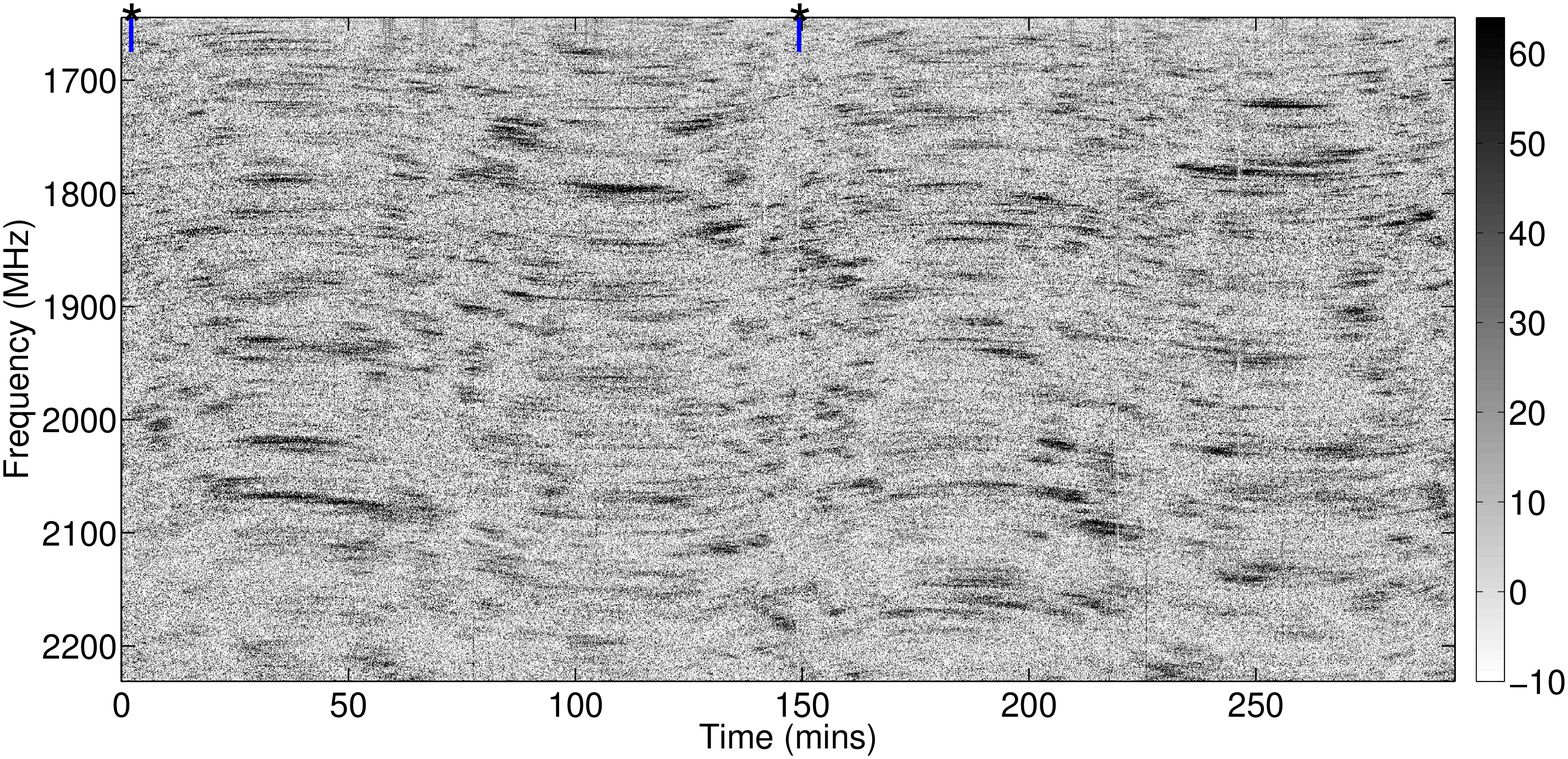} \\
\includegraphics[trim=55 0 0 35,clip,width=18cm]{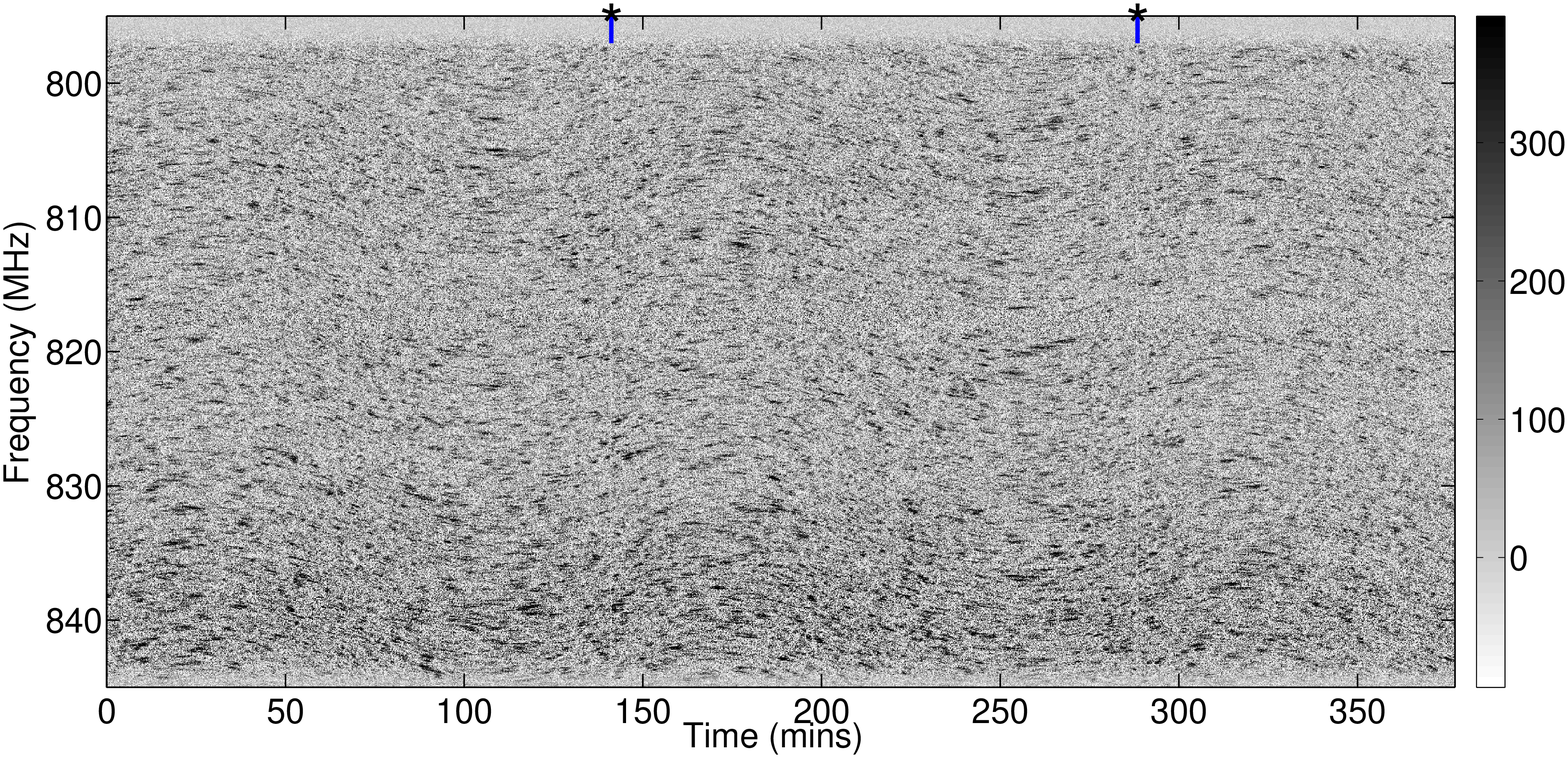}
\caption{Dynamic spectra of the A pulsar  taken at the 
Green Bank Telescope in 1024 channels at 10 s intervals. 
\it Upper: \rm MJD 53560 at 1900 MHz.
\it Lower: \rm MJD 53467 at 820 MHz.  The total durations differ
and a short line with a * on the top axis marks the time of A's eclipses.}
\label{fig:53560}
\end{figure*}

Figure \ref{fig:53560} shows the dynamic spectra of the A pulsar at
1900 MHz on MJD 53560 (top) and 820 MHz on MJD 53467 (bottom).  
The upper panel shows the mottled structure of ISS
with islands of high intensity (scintles) whose typical timescale 
varies from a few samples up to tens of samples, repeating over
the orbital period of 147 min. This observation should be interpreted as
the motion of the line of sight through a quasi-stationary spatial pattern
as the pulsar and the Earth move in their orbits and the IISM drifts in
a linear fashion. The time variation is determined by the spatial structure and the velocity
of the line of sight. The frequency correlation is only a few samples wide
and does not vary over the orbit, as discussed in \S\ref{sec:model0}.

The lower panel shows 2.5 orbital periods at 820 MHz.  There is minimal
RFI in this band and the eclipse by the B pulsar
can be seen as thin vertical lines of low intensities near 142 and 289 min. 
One can see that the ISS time scale is significantly shorter and the bandwidth is much
narrower, as expected for stronger scattering at the lower frequency. There are many more
scintles in this dynamic spectrum. This property 
and the absence of RFI make the analysis of the 820 MHz observations more
satisfactory than the 1900 MHz observations.
Sloping features, which are obvious in both plots, change sign over 
the orbital period. These are due to frequency dependent refraction as we
discuss in \S\ref{sec:refraction}. 

The dynamic spectra at 1900 MHz were often contaminated by RFI.
The full 800 MHz bandwidth was reduced to 600 MHz because of nearly continuous
RFI in the remainder of the band. In other cases RFI was
short-lived and we flagged regions of the dynamic spectra where RFI was
suspected and carried that flag array through subsequent processing. Flagged
data was simply excluded from all subsequent analysis. In Figure \ref{fig:53560}  
the flagged data were clipped at $\pm 3\sigma$ for plotting only.

\section{Characterization and Modelling of the Scintillation}
\label{sec:CharScint}

In many early ISS studies the dynamic spectrum was characterized by only 
two parameters - the characteristic widths in time and frequency.  These were usually
estimated from auto-covariance functions (acfs) versus time and frequency, in 
each case averaged over the other coordinate.  More generally the acf can be
computed in two dimensions with cuts along the two axes providing time and frequency widths. 
Recently observers have analyzed the secondary spectrum, which is the (2-dim) Fourier 
transform of the acf \citep{stinebring01}. For nearby strong young pulsars with highly anisotropic
scattering these secondary spectra show a wealth of interesting information in the form
of parabolic arcs. The secondary spectra of pulsar A do show parabolic arcs, but they are
not sufficiently well-defined to assist with our analysis. However we have found the 2-dim
acf very useful in estimating both the anisotropy of the spatial structure and the mean
phase gradient over the scattering disc. This is discussed in detail in section \S\ref{sec:refraction}.

\begin{figure}[htb]
\begin{tabular}{rl}
\includegraphics[ height=5.cm,trim=10 25 50 5, clip]{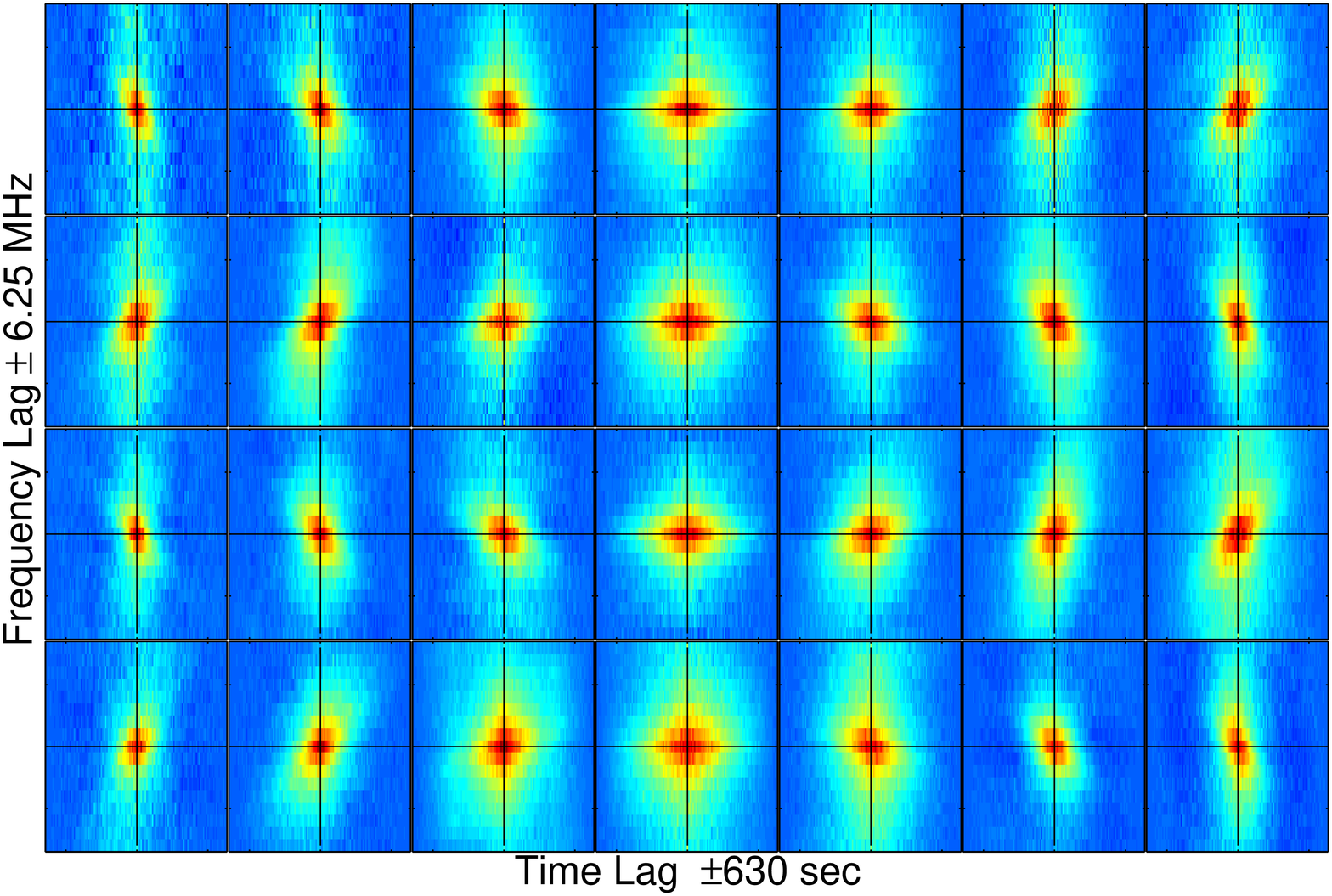} &  \includegraphics[height=4.8cm, trim=110 0 700 0, clip]{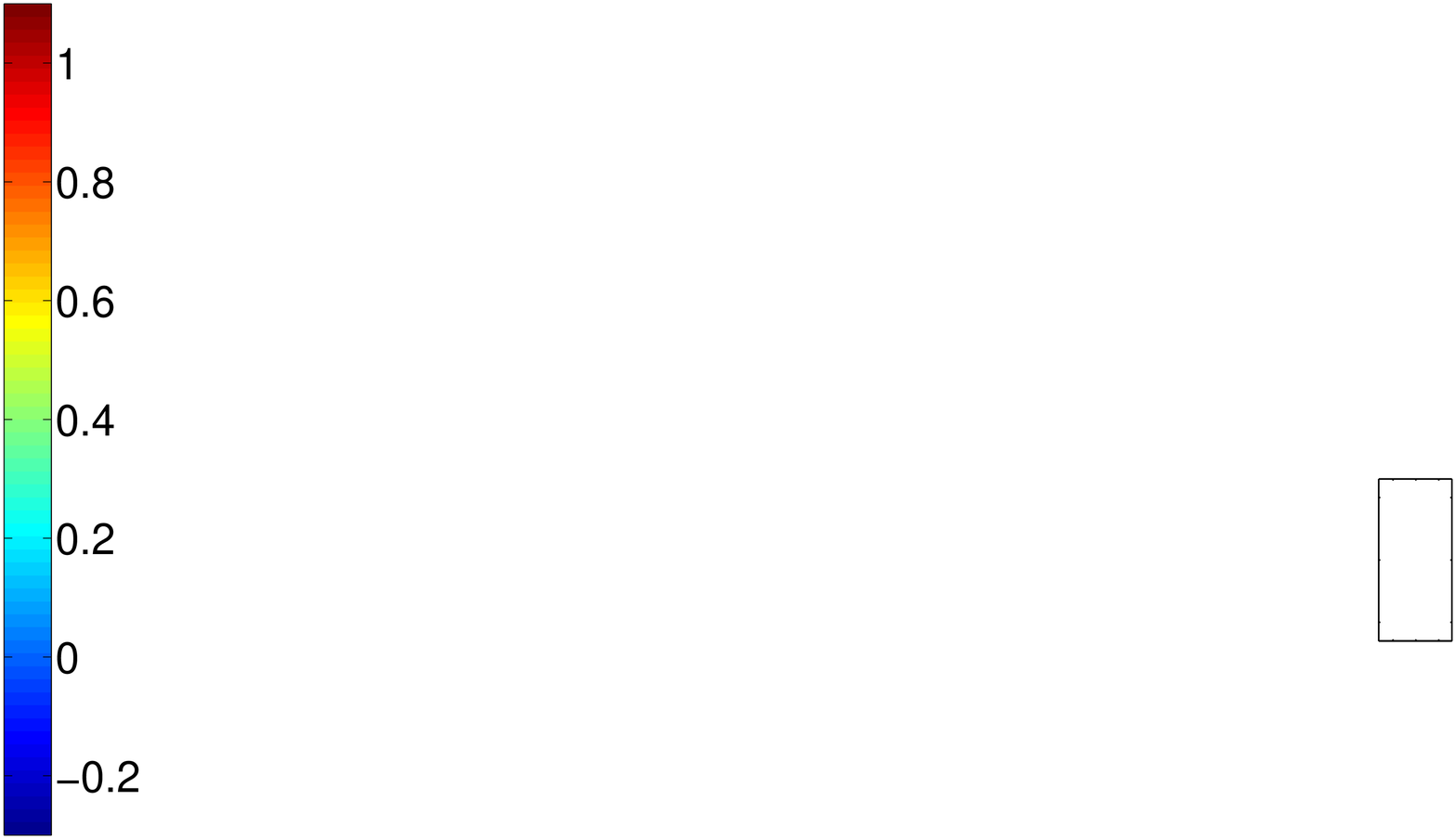}\\
\includegraphics[ height=5.cm,trim=10 25 50 5, clip]{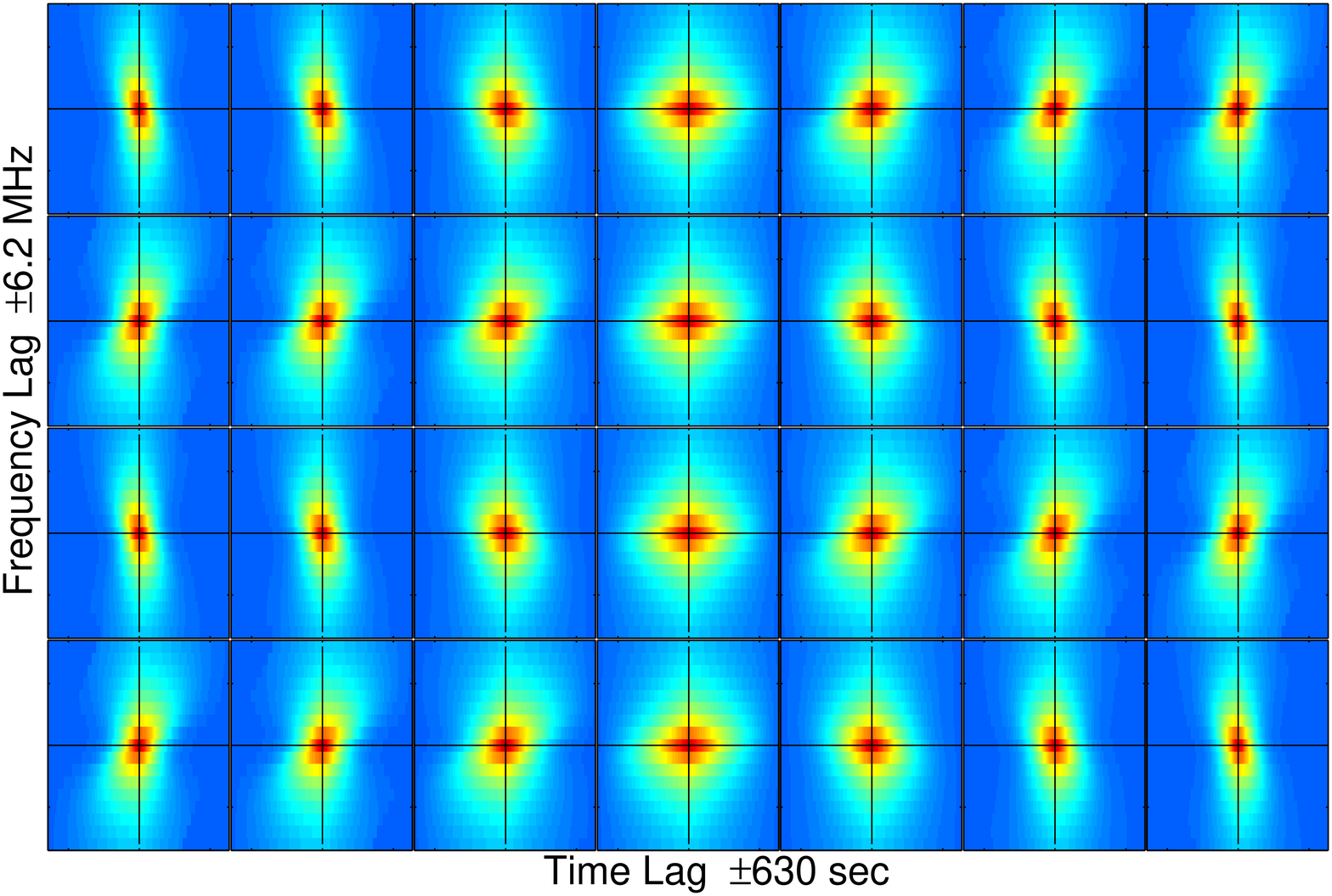} & \includegraphics [height=4.8cm, trim=110 0 700 0, clip]{colorbar.eps}\\
\end{tabular}
\caption{Frequency-time acfs for the dynamic spectrum  
in the top panel of figure \ref{fig:53560} (at 1900 MHz on MJD 53560).  {\it Upper} Observed;
{\it Lower} Model. Two complete orbital periods are shown, each divided into 14 blocks of 630s. 
Time advances left to right and top to bottom. The acfs are normalized by the variance of the ISS.
The noise spike at the origin is suppressed by the color table which saturates in dark red at unity.
Note the changing width in time lag, the 
nearly constant width in frequency lag and repeating patterns of 
positive and negative slope.}
\label{fig:ftacf}
\end{figure}

A theoretical model for the 2-dim acf $\rho_I(\tau,\nu)$ is developed in the Appendix. It is actually a ``cut''
through a 3-dim acf $\rho_I(\vec{r} = \vec{V} \tau,\nu)$ where $\vec{V}$ is the velocity of the line of sight
through the IISM and $\vec{r}$ is the transverse coordinate. 
Thus the apparent time scale is the spatial scale in the direction
of $\vec{V}$ divided by the speed. The width in frequency is inversely proportional to the
rms scattering angle and thus to the strength of scattering.
Here, as we want to study how the time and frequency scales of pulsar A vary 
over its orbit, we have computed the 2-dim acf
from short blocks of the dynamic spectrum, (which must be long enough to 
include at least one ISS timescale).   An example is shown in the upper panel
of Figure \ref{fig:ftacf} for the data from MJD 53560 shown in Figure \ref{fig:53560};
the lower panel is a best-fit theoretical model which we discuss in section \S\ref{sec:refraction}.

\subsection{ISS Characteristics and Model}
\label{sec:model0}

The characteristic scales in frequency and time have been defined by where
the auto-correlation functions fall to 0.5 and $e^{-1}$, respectively and we will
adhere to this convention.  In our
analysis we estimate these by fitting an ISS model to each acf, since
the fitting makes full use of the data.  
In deriving the model we assume the scintillations are dominated by scattering from
a thin region a distance $z_0$ from the observer. The electron density 
fluctuations in the screen are described by a Kolmogorov wavenumber spectrum which is homogeneous, 
at least over the scattering disc. The pulsar is at a distance of $z_p$ from the observer and the 
fractional distance from the pulsar to the screen is $s=(z_p-z_0)/z_p$.
Note that this thin scattering region can dominate the scattering, which is a path integral
over density squared, but may not dominate the dispersion, which is a path integral over density.
We also assume that refractive variations are negligible over the scales of interest,
i.e. the scintillations are in the diffractive limit.
This model is the starting point for many studies of ISS, which have been used to
investigate fine-scale turbulence in the interstellar plasma \citep{ars95, CL05}.
These studies assumed an isotropic density spectrum, but as recent observations such as \citet{brisken} have shown evidence for anisotropy, we also include it here. 

Propagation through such a layer causes a phase 
modulation which is usefully characterized by the phase structure function
$D_\phi (\bdma{\sigma}) = \langle (\phi_p (\bdma{r}) - \phi_p (\bdma{r+\sigma}))^2\rangle$,
where $\phi_p (\bdma{r})$ is the plasma phase contribution at transverse coordinate $\bdma{r}$. 
The electric field correlation at the output of the phase screen is
$\Gamma_E (\vsigma) = \exp(-0.5 D_{\phi}(\vsigma))$, and it is invariant with distance.

We describe the anisotropy by two quantities, 
the axial ratio $\ar$ and the orientation of the major axis $\psi_{AR}$ of the
inhomogeneities in the plasma density.  In terms of the major ($\sigma_{maj}$)
and minor axes ($\sigma_{min}$) the structure function can then be written
\be
D_{\phi}(\vsigma) &=&  \Bigl(\frac{\sigma_{maj}^2}{s_0^2 \ar} +  \frac{\ar \sigma_{min}^2}{s_0^2}\Bigr)^{5/6}.
\ee
\noindent In rotated coordinates ($\sigma_x,\sigma_y$) the quadratic form becomes
\be
D_{\phi}(\vsigma) &=& Q(\vsigma)^{5/6} s_0^{-5/3} \; , \nonumber \\
Q(\vsigma) &=& a \sigma_x^2 + b \sigma_y^2 + c \sigma_x\sigma_y \; ,
\label{eq:dphi}
\ee
\noindent 
with the major axis rotated by $\psi_{AR}$ clockwise from the x-axis. 
The mean diffractive scale $\siss$ decreases as the strength of scattering  
increases.   In this work we find it convenient to define anisotropy in terms of the
bounded parameter $R$,
\be
R = (\ar^2-1)/(\ar^2+1) 
\label{eq:RAR}
\ee
which lies in the range 0 to 1. The coefficients of the quadratic form $Q$ become:
\be
a &=& [1 - R \cos(2\psi_{AR})]/\sqrt{1-R^2} \nonumber  \\
b &=& [1 + R \cos(2\psi_{AR})]/\sqrt{1-R^2} \\
c &=& -2 R \sin(2\psi_{AR})/\sqrt{1-R^2} \nonumber 
\label{eq:abcR}
\ee
In the diffractive scintillation limit the spatial correlation function for
intensity is $C_I (\vsigma) = |\Gamma_E (\vsigma)|^2$ at the Earth. There are no
intensity fluctuations at the output of the screen. The temporal
correlation function at the Earth is $C_I (\tau) = \exp(-D_{\phi}(\bdma{V_{los}}\tau))$, 
where $\bdma{V_{los}} = (1-s)\bdma{V_{PA} } + s \bdma{V_E} - \bdma{V_{IS}}$ is the transverse velocity of 
the ``line of sight'' with respect to the plasma.
Here $V_E$, $V_{IS}$ and $V_{PA}$ are the  velocities of the Earth, the plasma, and pulsar A
respectively \citep{cr98}, all defined with respect to the Sun.

\begin{figure}[htb]
\begin{tabular}{ll}
\hspace{-3mm}\includegraphics[width=4.5 cm]{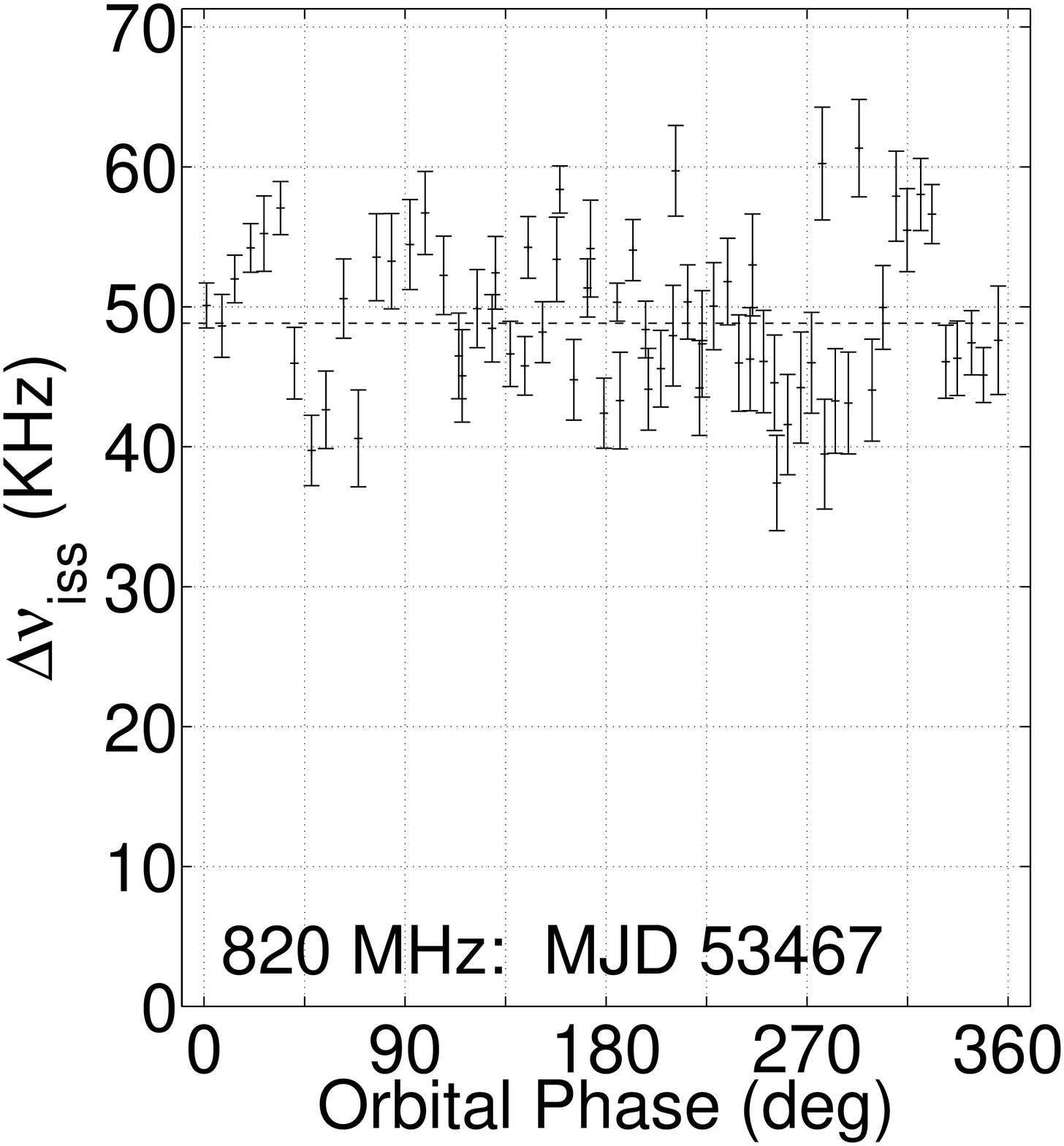} &  
\hspace{-6mm}\includegraphics[width=4.55 cm]{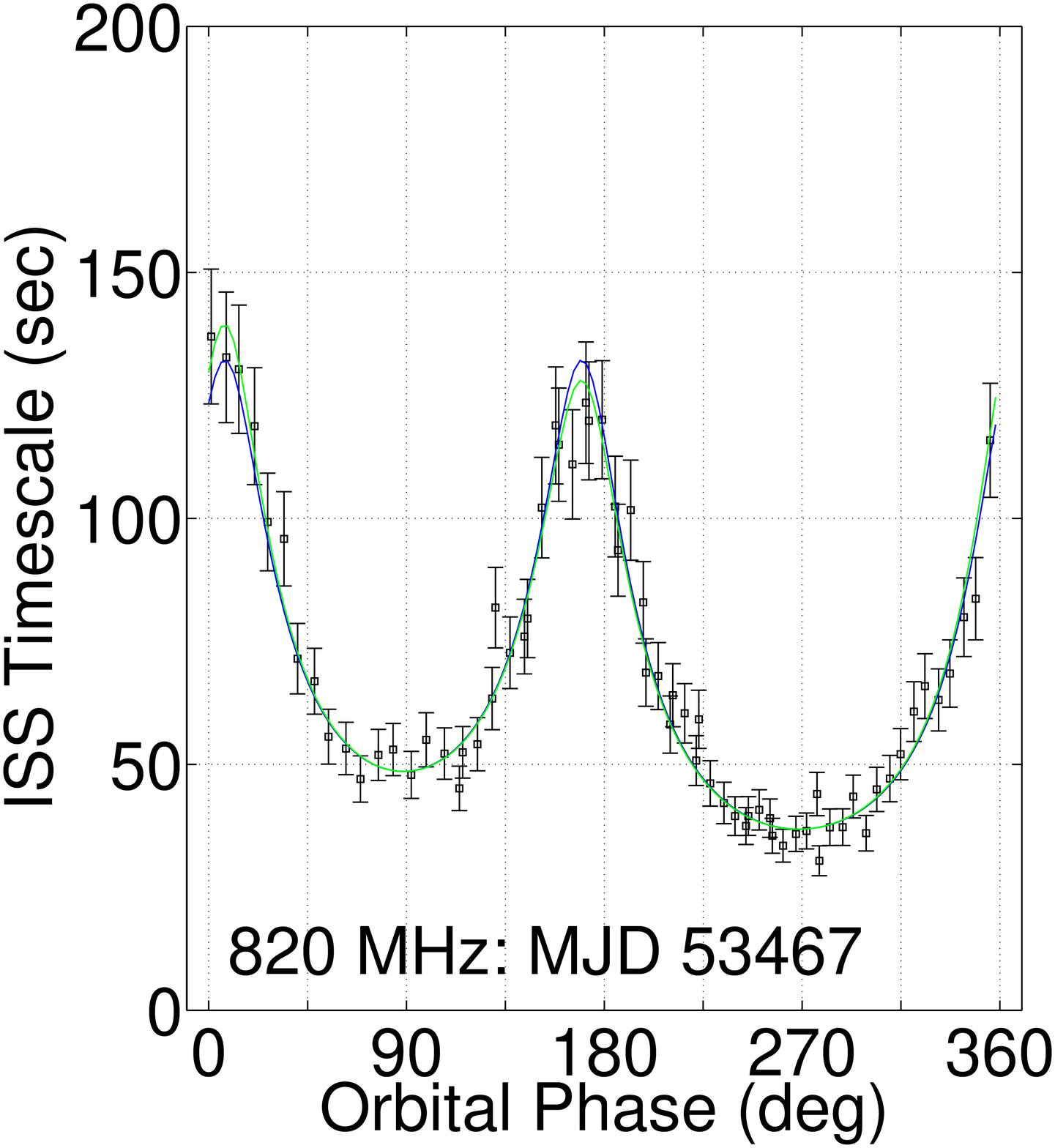} \\
\end{tabular}
\caption{Frequency $\Delta \nu_{iss}$  (left) and Time $\tiss$ (right) scales for the  
dynamic spectrum shown in the lower panel of Figure 1 versus true anomaly
orbital phase $\phi$, as defined in \S\ref{sec:model1}. 
The channel bandwidth is indicated by the dashed
line in the left panel. The theoretical best fit model is overplotted in the right panel. The error bars are 
$\pm1\sigma$.
}
\label{fig:deltanu_phi}
\end{figure}  

The acf versus frequency in the diffractive limit is more complex \citep{lov70}. It 
involves a Fourier-like integral which was solved by \citet{ar81} for the case of an
isotropic Kolmogorov spectrum.   
Here we have done this integral numerically for the anisotropic model and fitted it
to the acf of each block in each observing epoch. From this fit we obtain
the parameters $\Delta \nu_{iss}$ and $\tiss$ as a function of orbital phase $\phi$.
Details are given in Appendix \ref{app:ftacf}.  
The two scales $\Delta \nu_{iss}(\phi)$ and $\tiss(\phi)$ are plotted in Figure \ref{fig:deltanu_phi}
for the 820 MHz dynamic spectra shown in the lower panel of Figure 1. The model predicts that 
$\Delta \nu_{iss}$ should be independent of the pulsar velocity and 
indeed there is no discernible correlation with $\phi$, however
there is considerably more scatter in the estimates from each block than the 
fitting errors, as discussed by \citet{col10}.   As $\Delta \nu_{iss}$
is comparable to the channel bandwidth, we have barely resolved the ISS in frequency.
As expected $\tiss$ varies strongly over the binary orbit, primarily because of the variation in
the pulsar velocity.

The error bars on $\tiss$ are the formal standard errors of the fit of a covariance model to the measured
covariance. Most of the variance is due to white noise, so the error bars have not been corrected
for the correlation between measured covariance points. These error bars are used in a subsequent
weighted fit;  however, the error estimates resulting from this subsequent fit are independent of the scaling of the errors in $\tiss$.\footnote{Errors are 1 standard deviation throughout unless defined otherwise}

\subsection{ISS Timescale Variation over the Pulsar Orbit} 
\label{sec:model1}

The timescale $\tiss$ is easily determined in the isotropic case ($A_R$ = 1, $R = 0$),
to be $\tiss = s_0/V_{los}$. In the general case one must solve 
$\exp[-D_{\phi}(\bdma{V_{los}}\tiss)] = 1/e$ for $\tiss$:
\be
(1/\tiss)^2 &=& Q(\mathbf{V_{los}})/\siss^2 \; ,
\label{eq:tisssq1}
\ee
so here too $\tiss$ depends inversely on $V_{los}$. 

In our case pulsar A is in a binary system so that $\bdma{V_{PA}}$ 
is the sum of the binary center of mass velocity $\bdma{V_P}$ and A's orbital  velocity
$\bdma{V_{oA}}$ about the center of mass, 
which have been measured accurately from pulsar timing \citep{kramer06}.
It is the effect of the varying orbital velocity
that is responsible for the variation in $\tiss(\phi)$ over the orbital period
in Figure \ref{fig:deltanu_phi}. We
can then fit a model to $\tiss(\phi)$ and use the known $\bdma{V_{oA}}$ to
calibrate a model for the other variables ($s_0$, $s$, and $\bdma{V_{IS}}$).
This was first proposed by \citet{lyne84} and used by \citet{ord02} 
and \citet{ransom04} under isotropic scattering and generalized to anisotropic
scattering in paper 1.
The technique is most valuable for short period binaries 
because the orbit lies entirely within the scattering disc and a homogeneous 
scattering model fits the observations over a few orbits very well.
We will show that the time series $\tiss(\phi)$ possesses 5 degrees of freedom
and one could expect, in an isotropic plasma, to determine the scattering variables listed above
and also the orbital inclination $i$.
When the plasma is anisotropic there are two more unknowns in $A_R$ and
$\psi_{AR}$ and a complete solution requires more information.

As the orbital velocity of the pulsar is the reference, we use equation \ref{eq:tisssq1}
in the pulsar frame where the spatial scale is $s_p = s_0/(1-s)$.
The appropriate ``scintillation velocity'' is then
$\bdma{V_A} =\bdma{V_{los}}/(1 - s)  = \bdma{V_C} + \bdma{V_{oA}}$, where
\be
\bdma{V_C} = \bdma{V_{P}}+ \bdma{V_E} s/(1-s)  - \bdma{V_{IS}}/(1-s) 
\ee
which is constant during an observation, but obviously varies over the year.
We can write the transverse scintillation velocity of A $\bdma{V_A}$, 
in terms of $\bdma{V_C}$ and its true anomaly $\theta$ and its orbital phase from the line of nodes
$\phi = \omega + \theta$, where $\omega $ is the longitude of periastron, as follows:
\begin{eqnarray}
V_{Ax} & = & \vcx + V_o [ e \sin\theta \cos\phi - (1 + e \cos\theta) \sin\phi] \nonumber\\
      & = & \vcx - V_o e \sin\omega - V_o \sin\phi \nonumber \\
V_{Ay} & = & \vcy + \cos i [V_o  e \sin\theta \sin\phi + V_o( 1 + e \cos\theta)\cos\phi] \nonumber\\
     & = &  \vcy + \cos i (V_oe\cos\omega + V_o\cos\phi).
\label{eq:VxVy}
\end{eqnarray}
Here $(V_{Ax},V_{Ay})$ is in agreement with the equations  
given by \cite{ord02}, but differs from equations (4) and (5) of \citet{bogdanov}.  
Here $V_o$ is a mean orbital velocity given by
$V_o = 2\pi a/(P_{\rm b}\sqrt{1-e^2})$ in terms of 
$P_b$ the period, $a$ the semi-major axis of A and $e$ the eccentricity of
the orbit. 
We define the inclination $i$ as the angle between the orbital 
angular momentum $\mathbf{l}$ and the direction from the Earth toward the 
center of mass $\mathbf{\hat{s}}$.  We choose the $x$-coordinate 
along the line of nodes  ($\mathbf{\hat{x}}= \mathbf{\hat{s}\times\hat{l}}/\sin i$ 
and $\mathbf{\hat{y}}= \mathbf{\hat{s}\times\hat{x}}$), which with an inclination
near $90 \deg$ would make the angular momentum nearly
anti-parallel to the $y$-axis.

We rewrite equation \ref{eq:tisssq1} in the pulsar frame
and model the data $\tiss(\phi)$ by:
\begin{eqnarray}
(1/\tiss)^2 &=& Q(\mathbf{V_A})/\sisp^2 \nonumber \\
&=& (a V_{Ax}^2 + b V_{Ay}^2 + c V_{Ax} V_{Ay})/\sisp^2.
\label{eq:tissquad}
\end{eqnarray}
When the velocities from equation (\ref{eq:VxVy}) are substituted into equation (\ref{eq:tissquad}) 
we obtain the expression as a sum of five harmonics.
\begin{eqnarray}
(1/\tiss(\phi))^2 &=& K_0 + K_S \sin\phi + K_C \cos\phi \nonumber \\
&&+ K_{S2} \sin 2 \phi
 + K_{C2} \cos2 \phi ,
\label{eq:tissphi}
\end{eqnarray}
where the harmonic coefficients are shown below.
\begin{eqnarray}
K_0    & = & [0.5 V_o^2 ( a + b \cos^2 i) + a (\vcx - V_o e \sin \omega )^2 \nonumber \\ &&+
	b (\vcy + V_o e \cos \omega \cos i )^2 + \nonumber\\
       & & c (\vcx - V_o e \sin \omega )(\vcy + V_o e \cos\omega \cos i )]/\sisp^2 \nonumber \\
K_S    & = & -V_o [2 a (\vcx - V_o e \sin \omega ) \nonumber \\
	&&+  c (\vcy + V_o e \cos i \cos \omega)]/\sisp^2 \nonumber \\
K_C    & = &  V_o \cos i [ c (\vcx - V_o e \sin \omega ) \nonumber \\ &&+ 
	2 b (\vcy + V_o e \cos i \cos\omega)]/\sisp^2 \nonumber \\
K_{S2} & = & - 0.5 c V_o^2 \cos i /\sisp^2 \nonumber \\
K_{C2} & = &  0.5 V_o^2 (-a + b \cos^2 i )/\sisp^2 
\label{eq:harm5}
\end{eqnarray}
These 5 harmonics carry all the
information in the data set, i.e. they are ``sufficient
statistics'' for the ISS with a general (eccentric) orbit. 

\begin{figure*}[thb]
\begin{tabular}{rll}
\hspace{-1mm}\includegraphics[trim=7 22 67 55,clip, width=5cm]{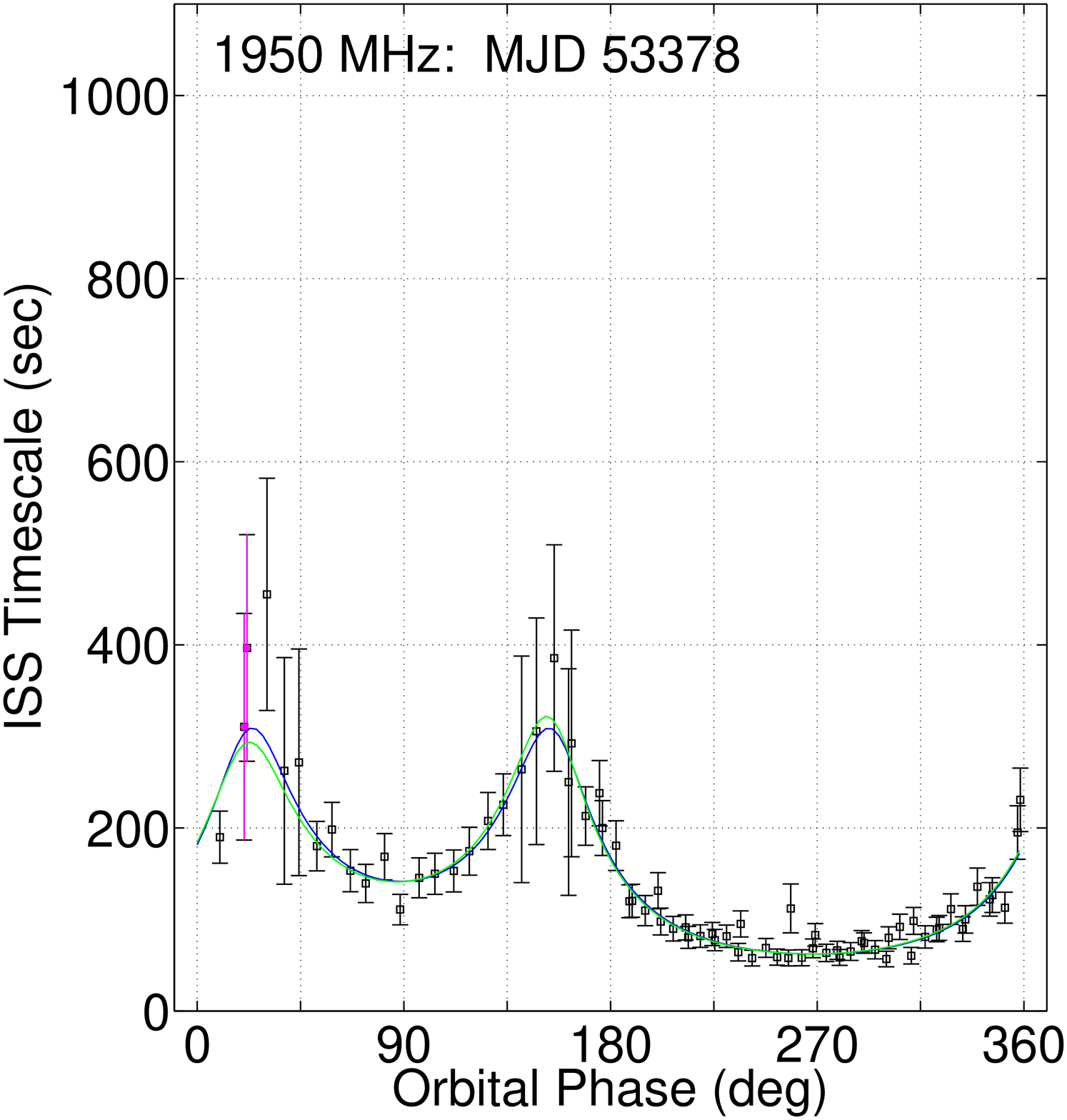} &
\hspace{-2mm}\includegraphics[trim=7 22 67 55,clip, width =5cm]{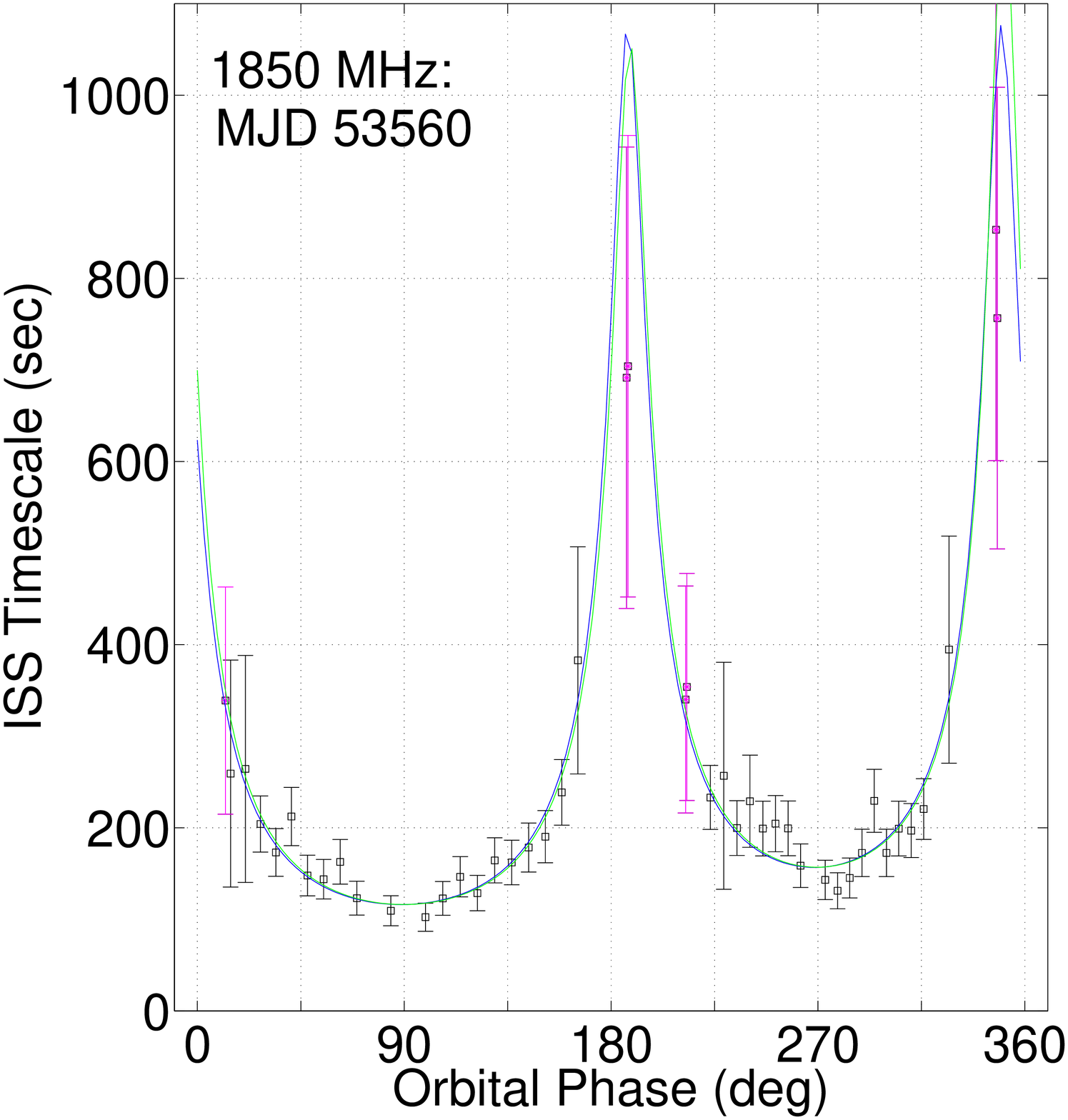} &
\hspace{-2mm}\includegraphics[trim=7 22 67 55,clip, width =5cm]{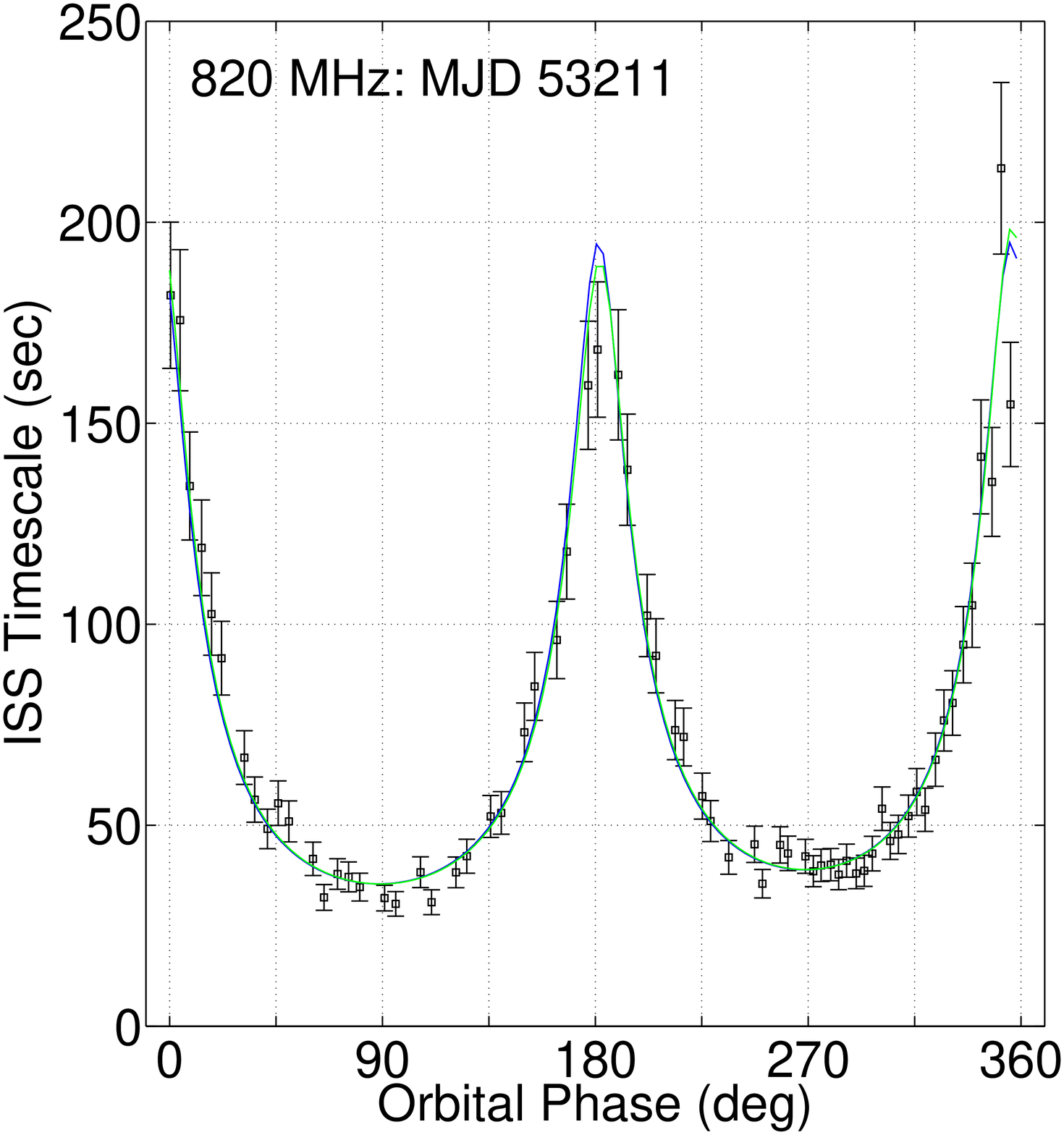} 
\end{tabular}
\caption{Plots of $\tiss(\phi)$ versus true anomaly orbital phase $\phi$
at 1900 MHz ({\it left to right}) from MJD 53378 \& 53560 and 
820 MHz from MJD 53211. Data are open symbols 
with error bars (magenta for the longer blocks). The 3 (5) parameter 
fits are blue (green) lines.}
\label{fig:tiss2}
\end{figure*}
The IISM can be characterized by five quantities ($\vcx, \vcy$, 
$\siss$, $R$ and $\psi_{AR}$), 
so that the 5 measurable parameters are insufficient to determine all 
the IISM parameters and $i$ unless $R$ is known to be zero. Further, the double pulsar is nearly 
edge-on so $K_{S2}$ and $K_C$ are small and the signal to noise ratio is marginal. 
One could still solve for $\siss, \vcx,$ and $\vcy$ if
$R$ and $\psi_{AR}$ were known.  
But since the ISS anisotropy is not 
known, extra information is needed as discussed in \S\ref{sec:earthmod}.
Thus we estimate the five harmonic coefficients as a first step in extracting
the full IISM parameter set and the inclination of the orbit.

In estimating $\tiss$ we split the dynamic
spectrum of A into time blocks of length 310 or 630~s,
subtracted the mean from each channel, 
and computed the temporal autocorrelation function $\rho(t)$ of each block
averaged over all channels.
We fitted a theoretical model to the autocorrelation of each block
beginning at the first non-zero time lag to avoid the ``noise spike''.
We used the theoretical form for diffractive ISS from a
Kolmogorov scattering medium $\rho(t) = \exp[-(t/\tiss)^{5/3}]$.
Since the time resolution in the dynamic spectrum is 5 or 10 sec 
we convolved the model acf with the appropriate triangular resolution function.
Also, since the block autocorrelation estimates
are weighted with a triangle function, we applied that same triangle 
weight to the model before fitting.  

In choosing the block length
there is a trade-off between the need for short block lengths, since the time
scale is continuously changing, and the need to keep the block length 
longer than the time scale.  
At 820 MHz $\tiss$ was always shorter than 310 s, and so we
we fixed the blocks at that length. 
However, at 1900 MHz the diffractive scintillation scale is larger and hence 
the longest time scales 
are sometimes greater than 310 s.  So at the higher frequency we computed
$\tiss$ for blocks of both 310 s and 630 s (1/14th of the orbital period).  
We then selected the estimates from 
the shorter blocks except where the longer block estimate gave $\tiss > 310$ s;
in which case we replaced the two nearest shorter blocks by the single block of 630 s.
We note that when $\tiss$ is large, it is quite sensitive to the form 
of the correlation function used in the model and is accompanied by larger errors.
This makes it difficult to exploit the fact that near the times of slow ISS
the scintillation velocity vector rotates rapidly and so has the potential
to determine the anisotropy of the ISS. 
At 1900 MHz we omitted the occasional blocks which were contaminated by 
RFI in all channels.

We then fitted the five harmonic model versus orbital phase $(\phi)$ to the estimates 
of $\tiss$ from each block.  Since equation (\ref{eq:tissphi}) is 
linear in the 5 harmonic coefficients, we started by fitting to 
$\tiss(\phi)^{-2}$, directly.  However, for a weighted fit we need 
errors in $\tiss^{-2}$, which could not be reliably obtained 
from the error ($\sigma_{\tiss}$) estimated for each $\tiss$.  
So we used the reciprocal of the square root of that equation to model 
$\tiss(\phi)$, after first smoothing it over the range of 
phases in each block.  We performed a non-linear least squares 
fit with residuals weighted by $1/\sigma_{\tiss}$ starting 
from the model parameters of the $\tiss^{-2}$ fit.
The fits for the 5 coefficients and their standard errors are listed in Table 
\ref{tab:harml}.  As expected the coefficients 
$\Kc$ and $\Kss$ are small for all epochs.  Examples of the fits are 
shown in figure \ref{fig:tiss2}.  The blue lines show
fits with only 3 harmonic coefficients ($\Kc=0, \Kss=0$) assuming $\cos i=0$, 
which are barely distinguishable from the fits with 5 coefficients.   

\begin{table}[!ht]
\caption{The five orbital harmonic coefficients versus date. The first 5 lines are 820 MHz and the latter 12
lines are at 1900 MHz. The units are $10^{-4}$ sec$^{-2}$. Day = MJD-53000.}
\label{tab:harml}
\smallskip
\begin{center}
\begin{tabular}{rrrrrr}
\tableline
\noalign{\smallskip}
Day &  \multicolumn{1}{c}{$K_0$} & \multicolumn{1}{c}{$K_s$} 
& \multicolumn{1}{c}{$K_c$} & \multicolumn{1}{c}{$K_{s2}$ }& 
\multicolumn{1}{c}{$K_{c2}$} \\
\tableline
$-3$&  3.46$\pm$.06 & $-2.63\pm$.08 & $-0.10 \pm$.03 &  0.10$\pm$.05 & $-2.77 \pm$.06  \\
211&  3.79$\pm$.10 &  $0.70 \pm$.11 & $-0.02 \pm$.02 & $-0.06 \pm$.07 & $-3.51 \pm$.10  \\
311&  4.76$\pm$.13 & $-3.63\pm$.16 & $-0.15 \pm$.05 &  0.17$\pm$.08 & $-4.02 \pm$.12  \\
379&  6.19$\pm$.14 & $-6.74\pm$.19 & $-0.04 \pm$.07 &  0.04$\pm$.08 & $-4.40 \pm$.12  \\
467&  3.25$\pm$.08 &$-1.57\pm$.10 & $-0.05 \pm$.03 &  0.01$\pm$.06 & $-2.57 \pm$.08  \\
\tableline
202&  0.96$\pm$.09 &  $0.41\pm$.15 &  $0.12\pm$.12 & $-0.10\pm$.09 & $-1.07\pm$.12  \\
203&  0.89$\pm$.05 &  $0.29\pm$.07 &  $0.00\pm$.01 &  $0.04\pm$.04 & $-0.86\pm$.05  \\
274&  0.92$\pm$.06 & $-0.40\pm$.08 &  $0.03\pm$.02 & $-0.06\pm$.04 & $-0.84\pm$.06  \\
312&  1.35$\pm$.06 & $-1.23\pm$.07 & $-0.04\pm$.02 &  $0.04\pm$.03 & $-1.15\pm$.06  \\
319&  1.41$\pm$.05 & $-1.39\pm$.07 & $-0.06\pm$.03 &  $0.05\pm$.04 & $-1.13\pm$.05  \\
374&  0.91$\pm$.05 & $-1.02\pm$.09 & $-0.02\pm$.04 &  $0.03\pm$.05 & $-0.59\pm$.06  \\
378&  0.94$\pm$.03 & $-1.05\pm$.05 & $-0.02\pm$.02 &  $0.02\pm$.03 & $-0.62\pm$.03  \\
415&  0.74$\pm$.04 & $-0.80\pm$.06 & $-0.00\pm$.03 &  $0.02\pm$.04 & $-0.53\pm$.04  \\
451&  0.84$\pm$.03 & $-0.71\pm$.05 & $-0.02\pm$.02 &  $0.05\pm$.03 & $-0.67\pm$.04  \\
462&  0.74$\pm$.02 & $-0.60\pm$.03 & $-0.04\pm$.01 &  $0.04\pm$.02 & $-0.58\pm$.02  \\
505&  0.54$\pm$.03 &  $0.06\pm$.04 & $-0.00\pm$.01 &  $0.01\pm$.03 & $-0.49\pm$.04  \\
560&  0.29$\pm$.01 &  $0.16\pm$.02 & $-0.00\pm$.00 & $-0.01\pm$.01 & $-0.26\pm$.01  \\
\\\tableline
\end{tabular}
\end{center}
\end{table}

\subsection{Annual Modulation by the Earth's Motion} 
\label{sec:earthmod}

\subsubsection{Analysis overview}

In Paper 1 we analyzed the observed variation in $\tiss$  over orbital phase at a single epoch
and found that any anisotropy in the ISS pattern has a strong affect on the estimate of the system velocity.
We also observed partial correlation in the ISS of the two pulsars.  In combination the observations 
could be fitted by any anisotropy $R>0.8$ which corresponded to
a very wide range of system velocities.  By assuming that $0.97\simgreat R \simgreat 0.8$ we constrained the system velocity
to be in the range 51 to 81 km s$^{-1}$, lower than 140 km s$^{-1}$ if $R=0$.   However the analysis did
not correct for the effects of the changing velocity of the Earth and any (unknown) velocity
of the interstellar scattering region.  Subsequent pulsar timing and VLBI  observations of the system
proper motion and parallax yielded a system velocity $\simless 20$ km s$^{-1}$.  In the
observations reported here we use observations over 18 months to resolve these difficulties. 

As noted in the previous section, our $\tiss(\phi)$ observations have been quantified by the
5 orbital harmonic coefficients measured at each epoch over the course of 18 months.
Our goal now is to fit a model to these coefficients assuming that the plasma parameters are 
constant over this period.  By including the (known) Earth's velocity we add two new unknowns, 
the location of the screen $s$ and the angle $\Omega$ defined CCW from celestial North through East
to the $x$-axis of the pulsar orbit. However the extra information in the annual variation can, in principle, 
be used to estimate the anisotropy. As the Earth's velocity provides an annual sinusoidal variation, the
responses of the 5 orbital harmonics also show annual and semiannual harmonics though they are
correlated. The annual-harmonics of the three strongest orbital-harmonics  $K_{C2}$, $K_S$ and $K_0$ 
provide 7 degrees of freedom and those of the two weaker orbital-harmonics $K_{S2}$ and $K_C$ add two 
more for a total of 9. Thus we can hope to estimate the five IISM quantities
($\vcx, \vcy$, $\siss$, $R$ and $\psi_{AR}$), plus $s$, $\Omega$ and $\cos i$.

We have found that we can obtain independent estimates of the anisotropy 
at each epoch by fitting the entire frequency-time acf, ($\rho_I(\tau,\nu)$) rather 
than just the time scale, but this can only be done for six of the best dynamic spectra, 
including all five observations at 820 MHz. The process is complicated by the 
need to also fit for a transverse gradient in the interstellar phase, because the phase gradient causes a first-order
effect on $\rho_I$ off the time axis. This process is discussed in section \S\ref{sec:refraction}. 
The result is that we are able
to estimate the five IISM parameters plus the phase gradient at each of these six epochs. This is very useful
in searching for time variations in parameters otherwise assumed to be constant.
 
Accordingly we use a hybrid procedure.  First we assume $\cos i = 0$ and fit the annual variation of
the three strong orbital harmonics. From this we can estimate the five IISM parameter plus $s$ and $\Omega$.
Second we include $\cos i$ in the parameters and add the two weaker orbital harmonics to the data to be fit.
This provides an estimate of $\cos i$ and also improves the estimate of the anisotropy. Then we fit the
frequency-time acfs for the 6 best dynamic spectra and obtain estimates of the time variation of the IISM
parameters. Finally we include these time variations in the fit of the annual variation of the five orbital harmonics. 

Note that the hybrid analysis is necessarily a compromise.  We are estimating diffractive scintillation
parameters from a few hours of data at each epoch, in a situation where we expect statistical 
variations in these parameters on the timescale for refractive scintillations (several days).   Thus the estimates 
are in the ``snapshot'' regime, as discussed by \citet{rnb}, who give theoretical predictions for the 
rms variation expected for various diffractive ISS parameters such as the anisotropy,
see also the simulations by \citet{col10}.    Thus an ideal model could allow
the ISS parameters to vary from one epoch to the next, as in the five observations at 820 MHz.
However, there is insufficient information to include such variations in the analysis at 1900 MHz
of $\tiss(\phi)$ and its harmonic coefficients.

\subsubsection{Non-stationarity in the Level of Turbulence}  

The harmonic coefficients $K_{C2}$ and $K_{S2}$ are independent of $V_E$ and should be
constant over the year. Although $K_{S2}$ is too weak to be useful in testing this hypothesis,
$K_{C2}$ is  accurately measured. It is inversely proportional to the square of
the spatial scale in the $x$-direction, i.e.~$\sisp^2 /{a}$ and to the frequency decorrelation width
$\Delta\nu_{iss}$.

\begin{figure}[htb]
\begin{tabular}{l}
\hspace{-3mm}\includegraphics[ width=7cm]{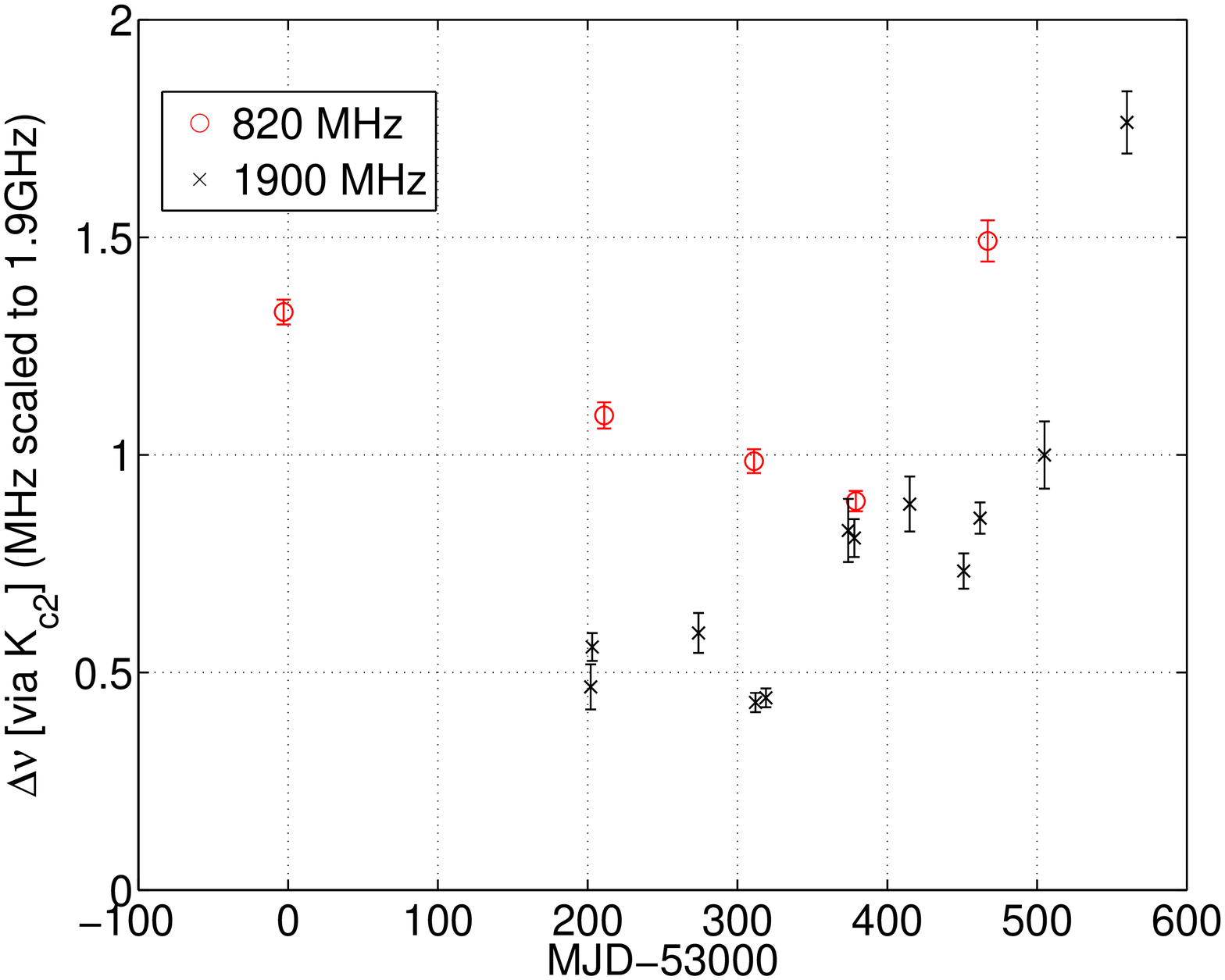} \\
\hspace{-3mm}\includegraphics[ width =7cm]{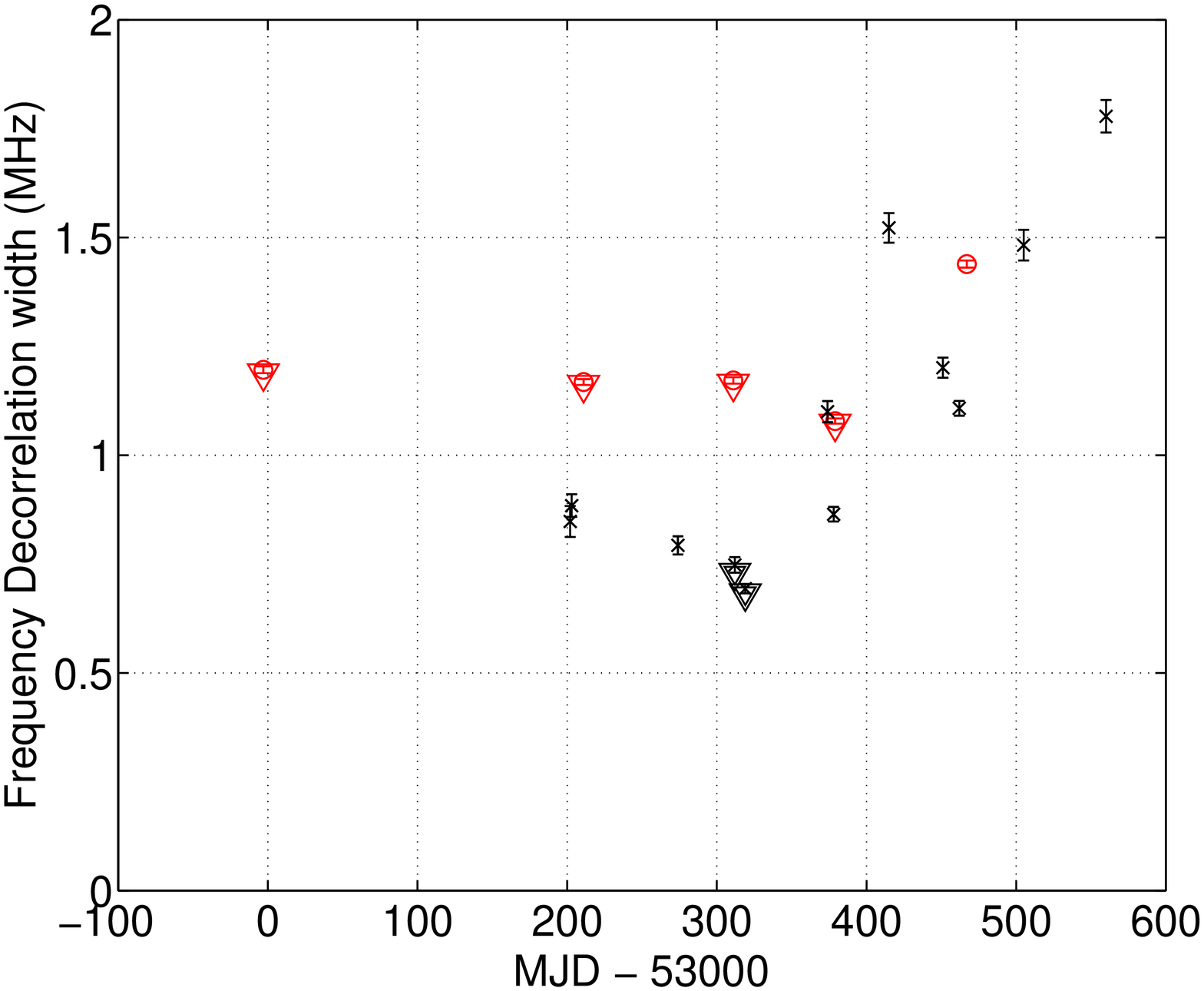} \\
\end{tabular}
\caption{Strength of ISS expressed as equivalent $\Delta\nu_{iss}$ at 1.9 GHz: 
{\it Upper} Derived from $K_{C2}$ and equation
\ref{eq:DelnuKC2}. {\it Lower} Estimated from fitting to acf (averaged over all
the blocks at each epoch). Error bars derived from the scatter
among the blocks at each epoch.}
\label{fig:delnu}
\end{figure}

We found that both $K_{C2}$ and $\Delta\nu_{iss}$ varied significantly over the 18 months of our 
observations. We have used $K_{C2}$ to infer $\Delta\nu_{iss}$ and plotted both the inferred and
directly measured $\Delta\nu_{iss}$ in Figure \ref{fig:delnu} (left and right panels respectively). The
equation used is 
\be
\Delta\nu_{KC2} = \frac{-a V_0^2}{2 K_{C2}}\frac{s}{1-s}\frac{2\pi\nu_m^2}{cL}
\label{eq:DelnuKC2}
\ee
where $L$ is the distance from the Earth to the pulsar and $\nu_m$ is the center frequency. One can see very
significant variations in both, although they are not identical. The difference is most probably because they have
different dependence on anisotropy. In particular only $K_{C2}$ depends on $\psi_{AR}$
and this parameter can vary between different realizations of the same random process.

We can remove the effect of these changes in $\sisp$ on the other parameters
by normalizing all the harmonic coefficients by $K_{C2}$ but this will not correct for minor
variations in the anisotropy.
This variation in the diffractive scale $\sisp$ seems greater than expected
from refractive effects, but  as in other pulsar observations it may indicate 
that the scattering medium is not well represented as homogeneous turbulence on a 
time scale of weeks to months \citep{hs08}.  Thus we might also expect to see some
variation in the anisotropy reflected in the harmonic analysis. 
We return to this question in \S\ref{sec:refraction}.

\subsubsection{Annual Variation in Harmonic Coefficients} 
\label{sec:annual}

\begin{figure}[ht]
\hspace{-2mm}\includegraphics[trim = 10 10 30 48,clip,width=9.cm]{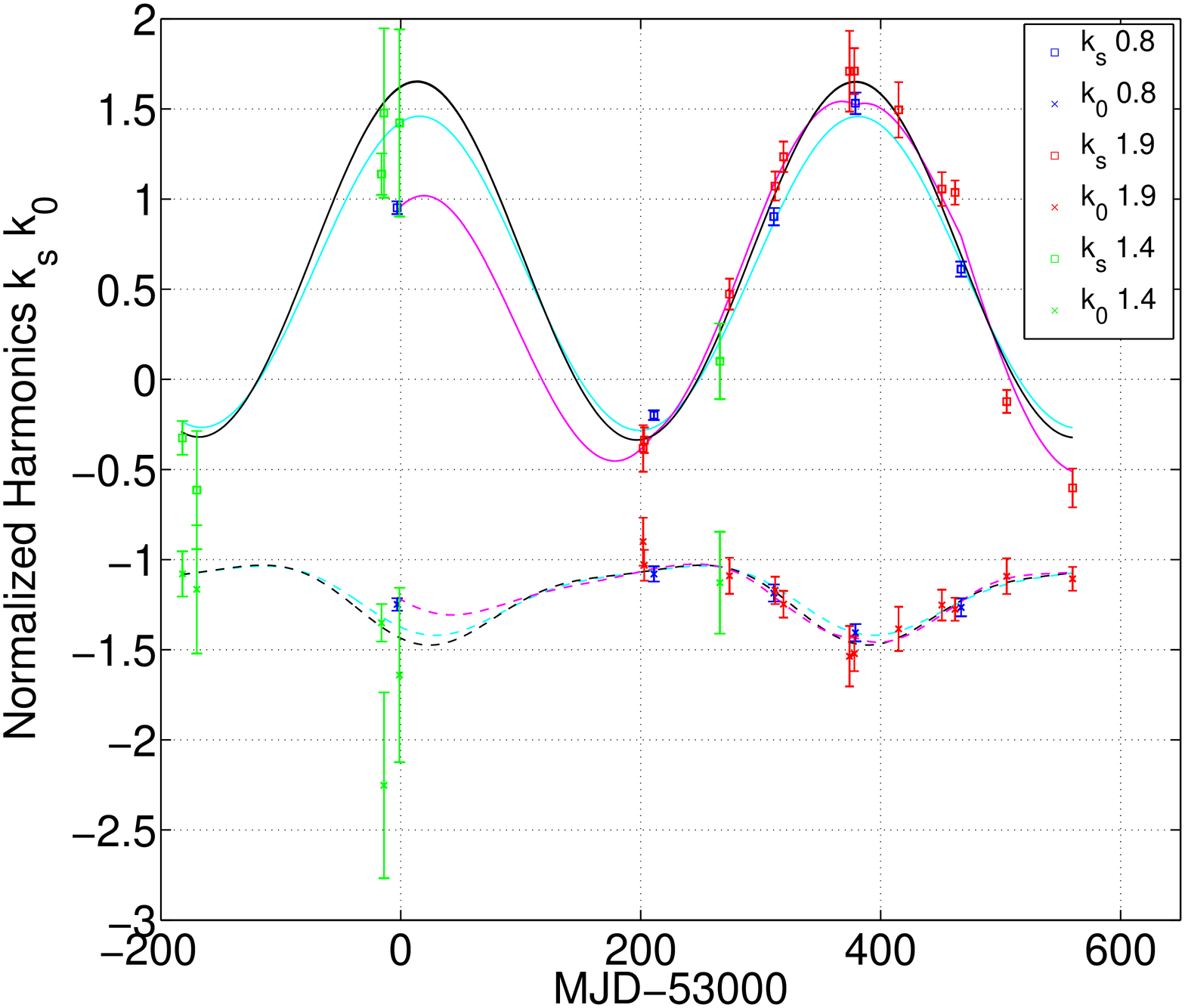}
\caption{Normalized harmonic coefficients 
$k_s$ and $k_0$ versus day. Symbols indicate the observing frequency in GHz.
Points at 1.4 GHz were not included in the fit, since their error bars were
substantially larger.  Three models are shown as
solid lines for $k_s$ and dashed lines for $k_0$; the cyan lines are constant velocity
fits including the point at day -3; the black lines are the same but excluding day -3.  
The magenta lines include day -3 but are fitted with variable velocity as discussed
in section 4.}
\label{fig:ksk0}
\end{figure}

The transverse velocity of the Earth and the proper motion of the pulsar are defined in celestial coordinates,
so we rotate them by the unknown angle $\Omega$ into (x,y) coordinates defined by the pulsar orbital plane,
i.e. $\bdma{V_{xy} = M V_{\alpha \delta}}$ where $\bdma{M}$ is
\be
\bdma{M}  =  \left( \begin{array}{rr}
\; \sin\Omega & \; \cos\Omega \\
 -\cos\Omega & \; \sin\Omega \end{array} \right).
\ee
\noindent
It remains convenient to define the unknown interstellar velocity in $x,y$ coordinates.
We then substitute $\bdma{V_C}$ into equation (\ref{eq:harm5})
and knowing the Earth's velocity obtain model equations for 
the harmonic coefficients at each epoch. 
We expect $\sisp$ to depend on the observing frequency, but we have
removed its influence by normalizing the two remaining 
coefficients by $\Kcc$, i.e. $k_s=\Ks/\Kcc$ and $k_0=\Ko/\Kcc$.
Thus we can combine the data from all 
frequency bands (820, 1400 and 1900 MHz)
by plotting $k_s$ and $k_0$ versus date in  figure \ref{fig:ksk0}. 
The overlap of values from different frequencies confirms the validity of this approach.

The equations for the two normalized coefficients versus date now involve 
$\bdma{V_C}$, $R$, $\psi_{AB}$, $s$, and $\Omega$, but it is convenient to
group them into the following three combinations:   
\be
u_x&=&(\vcx-eV_o\sin\omega)/V_o \;, \nonumber \\
\; u_y &=&\sqrt{b/a}(\vcy+eV_o\cos i\cos\omega)/V_o \\
w &=&c/\sqrt{ab} \nonumber
\label{eq:uxuyw}
\ee
then we have
\be
k_s &=& 4u_x + 2wu_y;   \nonumber \\
k_0 &=& -1 -2u_x^2 - 2w u_x u_y - 2 u_y^2   \; .
\label{eq:ksk0}
\ee
One can see that $k_s$ is linear in velocity, whereas $k_0$ is quadratic. The
known sinusoidal variation in $V_E$ will appear in $k_s$ scaled by $s/(1-s)$
and shifted in phase due to the rotation of coordinates by $\Omega$. So the
annual variation of $k_s$ has three degrees of freedom (including the constant).
The additional information provided by $k_0$ is the
square of a known sine wave plus an unknown constant. The semiannual harmonic
will provide an estimate of the axial ratio and the annual an estimate of $V_{ISy}$.
Thus including the constant term and $K_{C2}$ there are 7 degrees of freedom in total. This is
sufficient to estimate $s_0$, $s$, $\Omega$, $\bdma{V_{IS}}$, $R$ and $\psi_{AR}$. The
remaining two coefficients, $k_{s2}$ and $k_c$ are highly correlated and
add only two degrees of freedom.  However, with $\omega$ and/or $e$ known from timing
one can estimate $\cos i$.

In terms of the physical parameters we have:
\begin{equation}
u_x = p_1 + p_2 V_{Ex} \;\; ; \;\; u_y = p_3 + p_4 V_{Ey} ,
\label{eq:fg}
\end{equation}
where
\be 
p_1 &=& [V_{Px} - V_{ISx}/(1-s) -  e V_o\sin\omega ]/V_o \nonumber  \\ 
p_2 &=& s/[V_o(1-s)]  \nonumber \\ 
p_3 &=&  \sqrt{b/a}[V_{Py} - V_{ISy}/(1-s) + e V_o\cos i\cos\omega ]/V_o  \nonumber \\ 
p_4 &=&  p_2 \; \sqrt{b/a}.  \label{eq:p1-4}   \nonumber
\ee
In fitting the model we use the parameters obtained from the pulsar timing 
solution of \cite{kramer06} for eccentricity $e$ and
for  the longitude of periastron $\omega$ as a function of date; 
for the proper motion velocity we use: $\vpa=-17.8$, $\vpd=11.6$ km s$^{-1}$
from \citet{del09}.  This depends on the VLBI parallax distance of
$1.15^{+.22}_{-.16}$ kpc to the pulsar, which is larger
than 0.5 kpc based on the \citet{CL05} Galactic electron model.  
This smaller distance and the most recent proper motion estimated from
pulse timing (Kramer private communication 2014)
give a slower pulsar proper motion velocity: $\vpa=-5.3$, $\vpd=6.2$ km s$^{-1}$.   
In \S\ref{sec:ftacf} we discuss the small effect of using this lower pulsar velocity.

We initially optimized all 6 physical parameters
$s$, $\Omega$, $\bdma{V_{IS}}$, $R$ and $\psi_{AR}$
(under the assumption that $\cos i=0$)
to fit the data for $k_s$ and $k_0$ at 820 and 1900 MHz. The
result is shown in Figure \ref{fig:ksk0} by a cyan line. This fit
has a high reduced $\chi^2 \sim 9$, largely due to the obvious
outlier at day -3 (MJD 52997). This is an 820 MHz observation
with very small error bars but it is obviously discrepant because
it does not agree with a similar observation one year later. 
The dynamic spectrum for these data were at 5 s intervals, in
contrast to all the others which used 10 s intervals.  Though this
would change the pulse intensity and could influence $\tiss$, 
it should not distort the estimation of the harmonic coefficients in $\tiss$.
It suggests an error in our assumptions that the velocity and anisotropy
of the IISM were constant over this period, so we continued the
analysis with the data point at MJD 52997 excluded. The results are
shown as a blue line in Figure \ref{fig:ksk0} and tabulated in the first 
column of Table 2. The $\chi^2$ is reduced to a more reasonable 2.6,
but the errors on the velocity and the anisotropy are high.
We conclude that the location of
the scattering medium $s$ is quite accurately estimated, the orientation
of the pulsar orbit in celestial coordinates $\Omega$ is adequately determined, 
but the parameters of the IISM are weakly constrained.

\subsubsection{Constraining the orbital inclination and scattering anisotropy}
\label{sec:inclin}

\begin{figure}[htb]
\includegraphics[trim = 10 10 5 20, clip,width=8.cm]{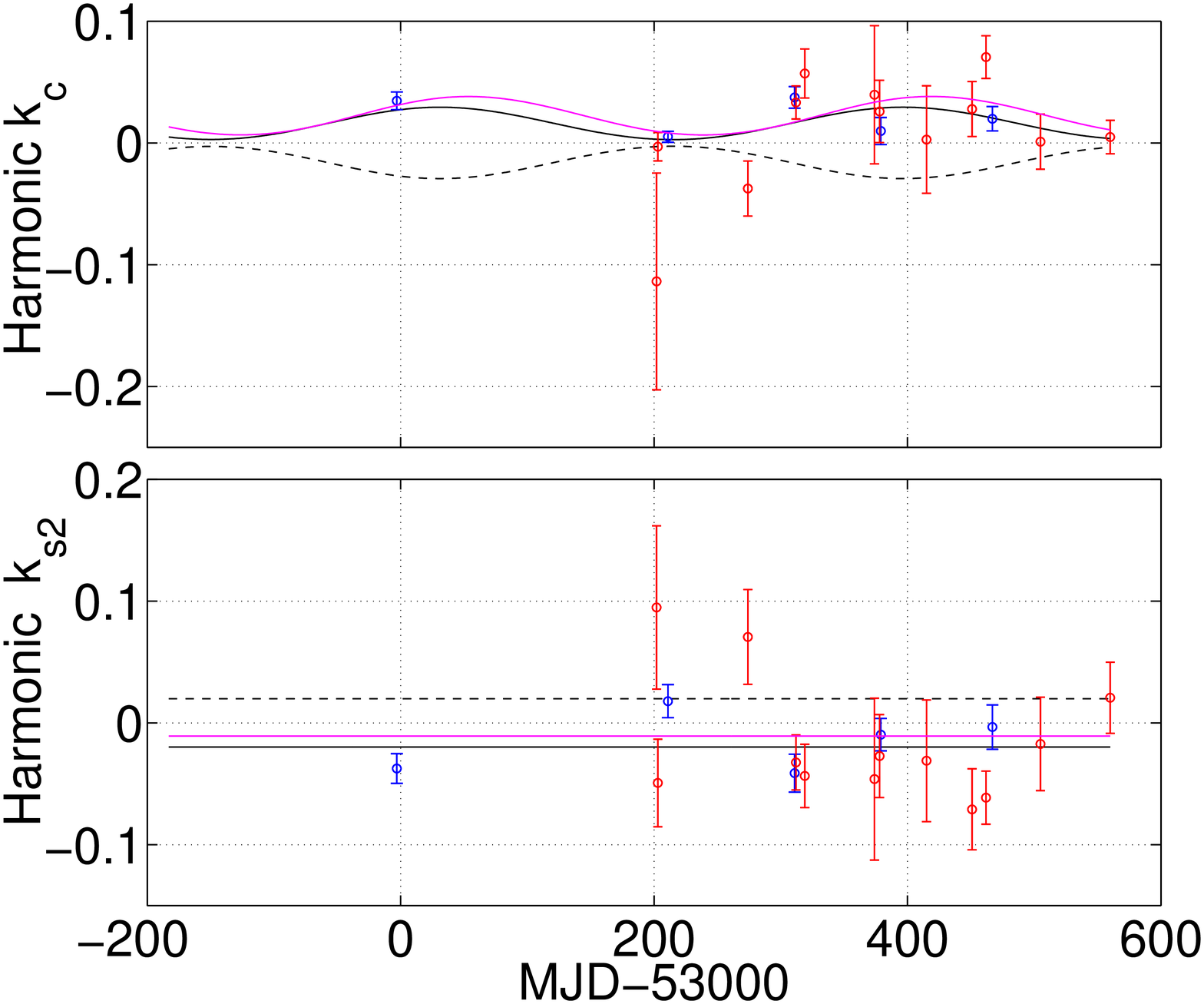}   
\caption{Normalized harmonic coefficients 
$k_c$ and $k_{s2}$ versus day (blue 820 MHz and red 1900 MHz).
Models were fit to all 4 harmonic coefficients, excluding those on day -3.
 The solid black curves are constant velocity models with
$i=88.7\deg$ and the dashed black curves are the same with $i=91.3\deg$. The
magenta curve is from a variable velocity fit discussed in section 4. }
\label{fig:kcks2}
\end{figure}

In view of the difficulties in fitting a fixed set of parameters to $k_s, k_0$ and with the hope of
determining the sign of $\cos i$, we tried fitting all four normalized harmonic coefficients with $\cos i$ 
restricted to $i=88.7\deg$ or $91.3\deg$.
The measured Shapiro delay \citep{kramer06} provides estimates of $\sin i$,  
leaving an ambiguity in the sign of $\cos i$, bounded by $|\cos i| = 0.023^{+.013}_{-.009}$.
The observed harmonics $k_c, k_{s2}$ are plotted Figure \ref{fig:kcks2}   
versus date and the best fit models for these two inclinations are over plotted. 
Even with the relatively large errors in these coefficients, it is clear that the solid curves for
$i = 88.7\deg$ fit better than the dashed curves for $i = 91.3\deg$. These two fits are
tabulated in the third and fourth columns of Table 2. It is remarkable how much the
inclusion of the fit for $\cos i$ and the two extra harmonics improved the error bars on
all the fitted parameters. The results for $i = 91.3\deg$ have a higher $\chi^2$, 
the parameters do not match those of the two harmonic fit well, and the IISM velocities
are well outside of the expected range. We conclude that the location of the screen and
the orientation of the orbital plane are now accurately estimated, and the estimates of the
anisotropy and velocity of the IISM are now useful. 
Encouraged by this result we included
$\cos i$ as a fitted parameter, obtaining $i=88.6\pm0.4\deg$.

\begin{table}[htb]
\caption{Parameters estimated from fitting to annual variation of
the normalized orbital harmonic coefficients at 0.8 \& 1.9 GHz,
excluding MJD 52997. The first column is a two harmonic fit, the others are four harmonic fits.
}
\label{tab:6params}
\smallskip
\begin{center}
\begin{tabular}{rrrr}
\tableline
\noalign{\smallskip}
Parameter & $i=90\deg$  & $i=91.3\deg$    &  $i=88.7\deg$   \\
\tableline   
$s$ & $0.71\pm 0.03$  & $0.70\pm.02$ & $0.70\pm 0.02$ \\
$\Omega$ (deg) & $69\pm31$ & $111\pm8$ & $61\pm8$ \\
$R$  & $0.76\pm.27$  & $0.96\pm.11$ & $0.71\pm0.21$ \\
$\psi_{AR}$ (deg) &  $72\pm36$  & $118\pm8$ & $61\pm9$ \\
$V_{ISx}$ (km s$^{-1}$) & $ -12\pm29$  & $-79\pm68$ & $ -9\pm11$ \\
$V_{ISy}$ (km s$^{-1}$) & $ 50\pm32$ & $\ge 100$ & $42\pm22$ \\
$N_{\rm dof}$ & 26  & 58 & 58 \\
Reduced $\chi^2$& 2.6 & 3.2 & 2.8 \\
\tableline
\end{tabular}
\end{center}
\end{table}

\begin{figure}[htb]
\center{
\begin{tabular}{ccc}
\hspace{-2mm}\includegraphics[trim = 45 35 35 35, clip,width=2.7cm]{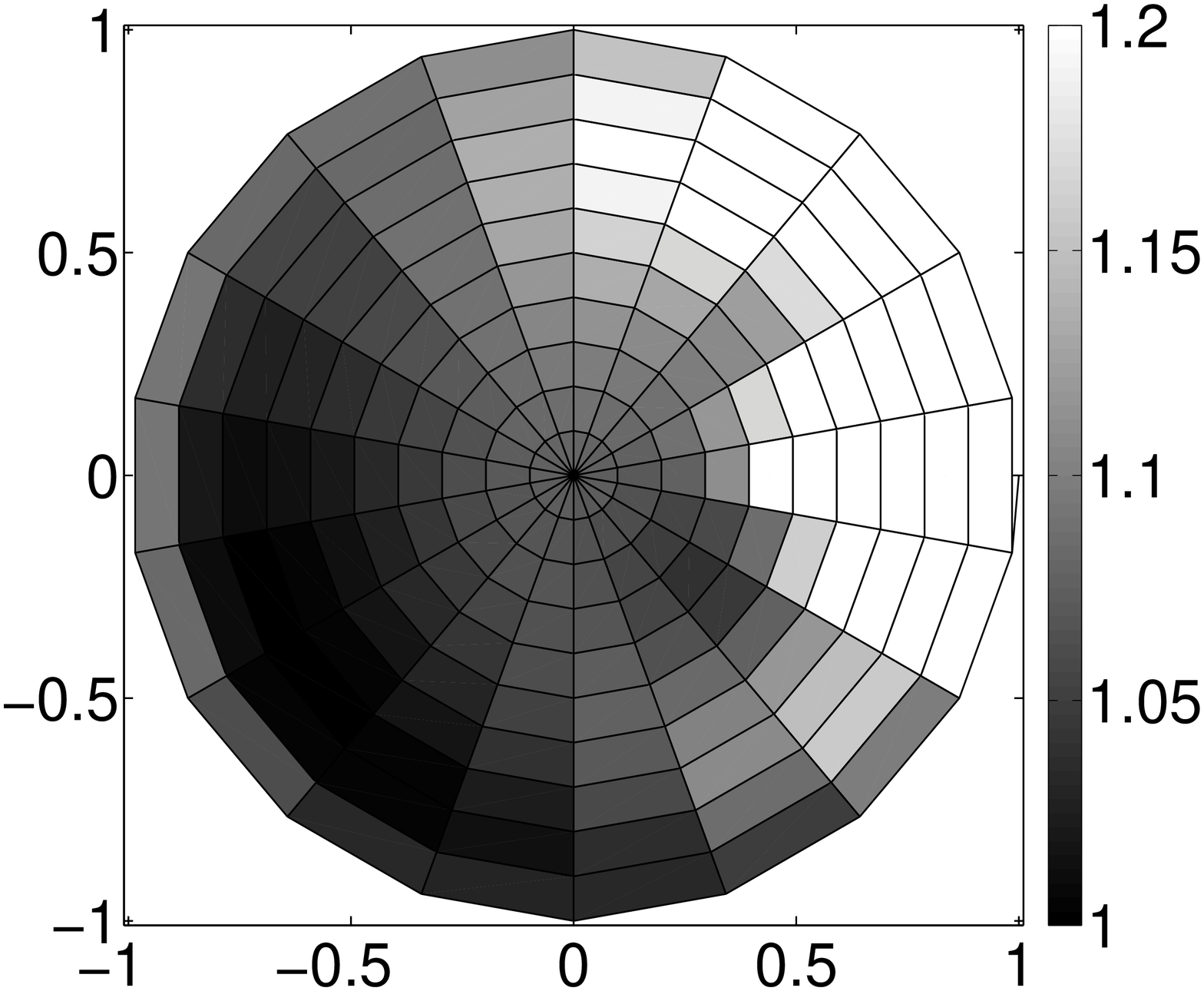} &
\hspace{-2mm}\includegraphics[trim = 45 35 35 35, clip,width=2.7cm]{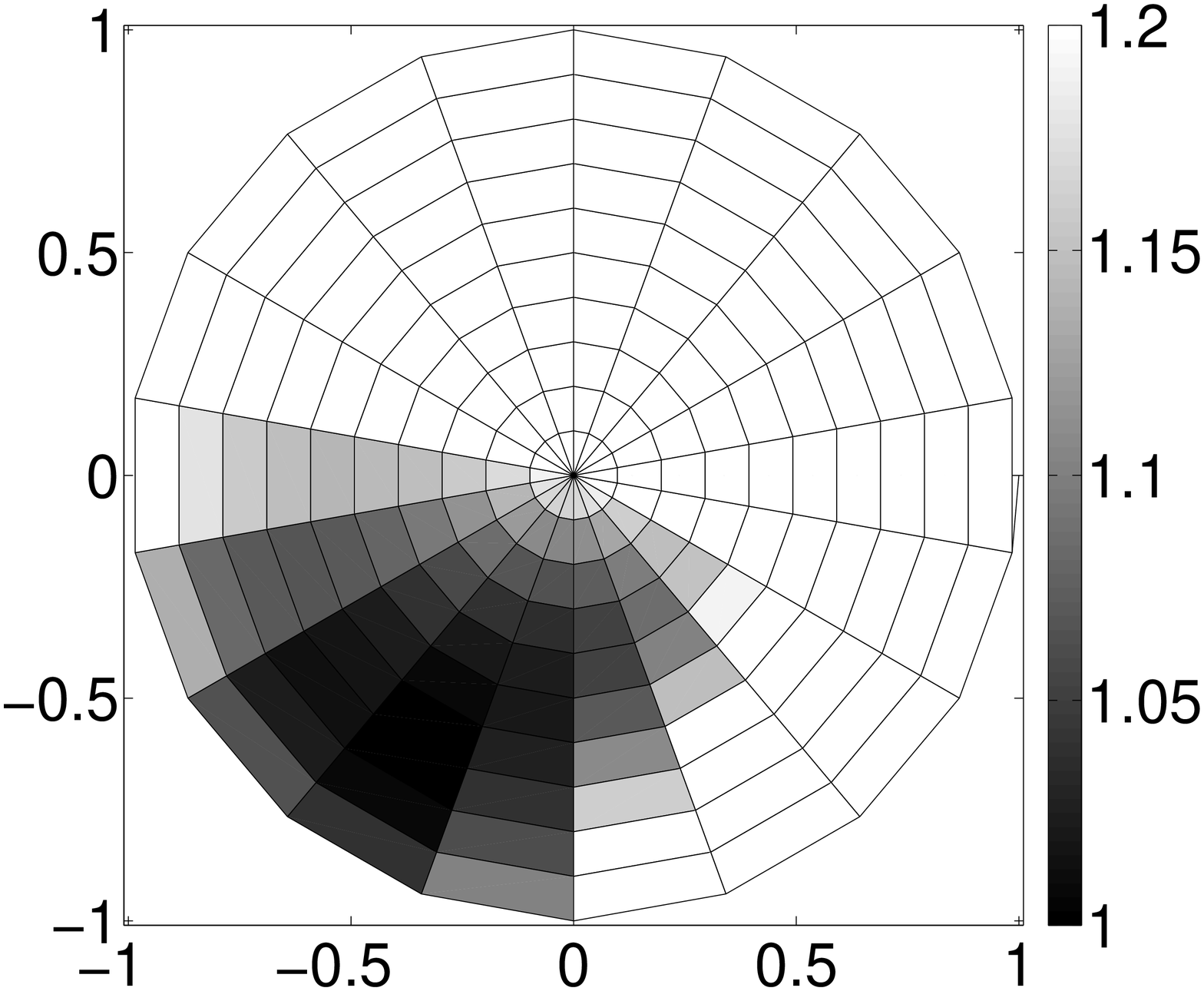} &
\hspace{-2mm}\includegraphics[trim = 45 35 35 35, clip,width=2.7cm]{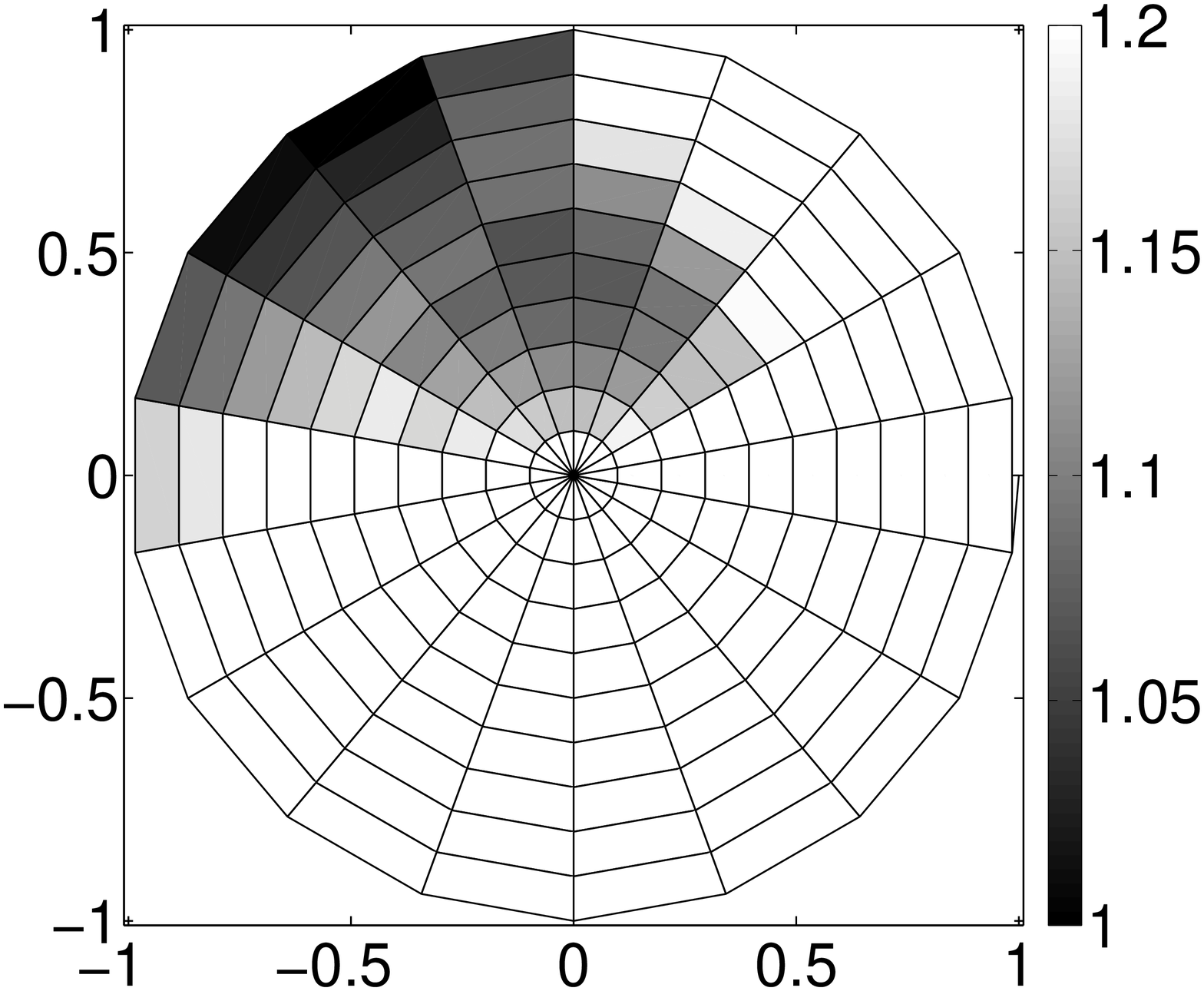} \\
\end{tabular}
}
\caption{Residuals ($\chi^2$) from fitting harmonic coefficients versus date 
(excluding MJD 52997) in a ``Poincar\'{e}'' polar plot of anisotropy ($R$, $2\psi_{AR}$).  
Residuals are normalized to the global minimum. 
\it Left: \rm Fitting 2 coefficients $k_s, k_0$. 
\it Center \rm Fitting 4 coefficients $k_s, k_0, k_c, k_{s2}$ with $i = 88.7\deg$.
\it Right \rm Fitting 4 coefficients $k_s, k_0, k_c, k_{s2}$ with $i = 91.3\deg$.
Looking toward the pulsars their orbit line of nodes ($x$-axis) is to the right in the plot.  
The orientation of the major axis ($\psi_{AR}$) is defined clockwise from $x$.   
}
\label{fig:tiss_wheel}
\end{figure}

To show the improvement in fitting four harmonics we redo the three fits discussed, i.e. the fit
for two harmonics with $\cos i = 0$ and the fits for four harmonics with $\cos i = \pm 0.023$.
Here we step $R$ and $2\psi_{AR}$ over the ranges: $0 \le R \le1$; $0 \le 2\psi_{AR} \le 2\pi$
while fitting for the other 4 parameters at each grid point. 
The left panel of figure \ref{fig:tiss_wheel} shows the sum of the squares of the 
residuals as a polar plot in this space, which could be called a 
Poincar\'{e} circle; see also a similar plot by \citet{grall97}.  

The left panel shows the result for the two harmonic fit with $i = 90\deg$. The mean squared error surface is
very broad, covering the left half of the Poincar\'{e} circle. The middle panel is for a four harmonic
fit with $i = 88.7\deg$. It is much more compact, but consistent with the left panel. The right panel is for
a four harmonic fit with $i = 91.3\deg$. It is completely disjoint with the other panels, and shows a much higher
axial ratio.  We conclude that $\cos i > 0$ and that the six parameters are
reasonably well constrained.

The fit is not completely satisfactory because we have not explained the discrepancies at day -3 and
day 560. It seems most likely that, in addition to the level of turbulence, which we know to vary, that the
velocity of the IISM or its anisotropy must also vary with time.
In searching for time variations, we need to obtain more information at each epoch because the three
primary harmonic coefficients are only capable of constraining $\sisp$ and two other parameters. However
we can assume that we know $s$, $\Omega$ and $\cos i$. So we need only constrain two more parameters
to obtain an independent fit at each epoch. We will show in the following section that this can be done using
the full 2-dim frequency-time acf.

\section{The frequency-time structure and large scale phase gradients}
\label{sec:refraction}

The dynamic spectra in figure \ref{fig:53560} show striking features which are tilted 
in the frequency-time plane. Such tilted structures have often been seen  in 
pulsar ISS observations \citep{r1, r2} and also in simulations of pulsar scattering \citep{col10}.
They are due to dispersive refraction in the interstellar medium, which causes a frequency dependent
spatial shift in the ISS pattern that is mapped to a frequency dependent time shift by the pulsar velocity.
The effect can be seen very clearly in upper panel of Figure \ref{fig:ftacf} where the observed
frequency-time (2-dim) acfs show regular changes in slope which are synchronous with the orbital period.
We find that this variation can be modeled well with a constant phase gradient through which the line of sight
moves during the pulsar orbit. The modeled acfs are shown in the right panel and are 
discussed below.

We computed 2-dim acfs for all the observing epochs at both 820 and 1900 MHz.
For the 820 MHz observations we were also able to fit models 
to the 2-dim acfs in all the blocks in each of the 5 epochs, 
providing estimates for the velocity, anisotropy, spatial scale,
and the phase gradient for those five epochs.  The block length was
320 sec for MJD 52997, which was sampled at 5 s intervals; for the 4 remaining epochs
the sample interval was $\approx10$ s and the blocks were 63 samples. 
At 1900 MHz the blocks were 630 s, but the fitting was much less successful
due to sporadic narrow band RFI and only the observations 
on MJD 53560 shown in Figures 1 and 2 provided a useful set of acfs.

The computation of the 2-dim acf from each block of data is such that it is tapered by a triangle 
in time lag but is essentially unbiassed versus frequency lag, because the recorded bandwidth
is much wider than that of the ISS. In addition the acf
contains a contribution from white noise as a spike at the origin. As described in \S\ref{sec:model1}
we fit a theoretical model to the temporal acf, which is well sampled, beginning 
at the first time lag. Then we extrapolate the model back to 
zero lag to estimate the variance of the ISS.  This is used both to replace the acf
at zero lag and to normalize the acfs at all frequency and time lags.
Correct normalization is essential in obtaining a good fit to the frequency 
variation because the frequency axis is rather 
under sampled and the point at the origin is very important in the fit.
We computed the theoretical model of the acf 
with higher resolution than the observations and filtered it to include the effect of 
finite resolutions in time and frequency and also multiplied it by
the triangular taper in time lag.

\subsection{Model Fitting of the  frequency-time acf}
\label{sec:ftacf}

As noted in \S\ref{sec:model0}, the spatial correlation of intensity 
$C_I (\vsigma) = \exp(-D_{\phi}(\vsigma s))$.  However, $C_{I}(\vsigma,\Delta \nu)$
is more complex as described in Appendix \ref{app:ftacf}.
The 2-dim (frequency-time) acf of a dynamic spectrum is $C_I (\vsigma=\bdma{V}_{los}\tau,\Delta\nu)$.
If $\bdma{V}_{los}$ is constant this 2-dim acf carries no information
on spatial anisotropy, but in a binary orbit $\bdma{V}_{los}$ varies over a considerable angular
range and it becomes possible to estimate both the anisotropy and the velocity.
However, we must also consider the effects of a transverse gradient in $\phi_p$ 
which can have a significant effect on $C_{I}(\vsigma,\Delta \nu)$,
as commonly seen in the ISS of some pulsars. Such a gradient causes refraction by an angle:
\be  
\bdma{\theta_{\rm p}} =  \bdma{\nabla}\phi_{\rm p}/k  \propto \nu^{-2}.
\label{eq:thetap}
\ee
The refraction displaces the ISS pattern by a transverse vector $\bdma{\sigma}_p$
which, due to dispersion, is $\propto \nu^{-2}$.    
In Appendices \ref{app:ftacf} \& \ref{app:ref} we include this refractive shift, assumed to be
constant over the pulsar orbit, in the theoretical model.   

The model involved the following 8 parameters:
$\bdma{\nabla}\phi_{\rm p}$,  characterized by $\vsigma_p$ relative to the
orbital $x$ axis; $R$, and $\psi_{AR}$; $\Delta\nu_{iss}$ and the time scaling factor 
$V_o/\siss$; and the normalized velocities $u_x, u_y$, defined in equation (\ref{eq:uxuyw}).
The fitting is constrained by the harmonic coefficients $k_s$ and $k_0$  on the date in question. As
$s$ and $\Omega$ are known, and $R$ and $\psi_{AR}$ are fitting parameters, we can use 
$k_s$ and $k_0$ to determine $u_x$ and $u_y$. Thus only 6 parameters need be fit.
However there is a sign ambiguity in $u_y$, which we resolved by 
matching its sign to that of the annual variation model shown in 
figure \ref{fig:ksk0} on that date.

\begin{figure}[tb]
\begin{tabular}{cl}
\includegraphics[height=4.5cm, clip]{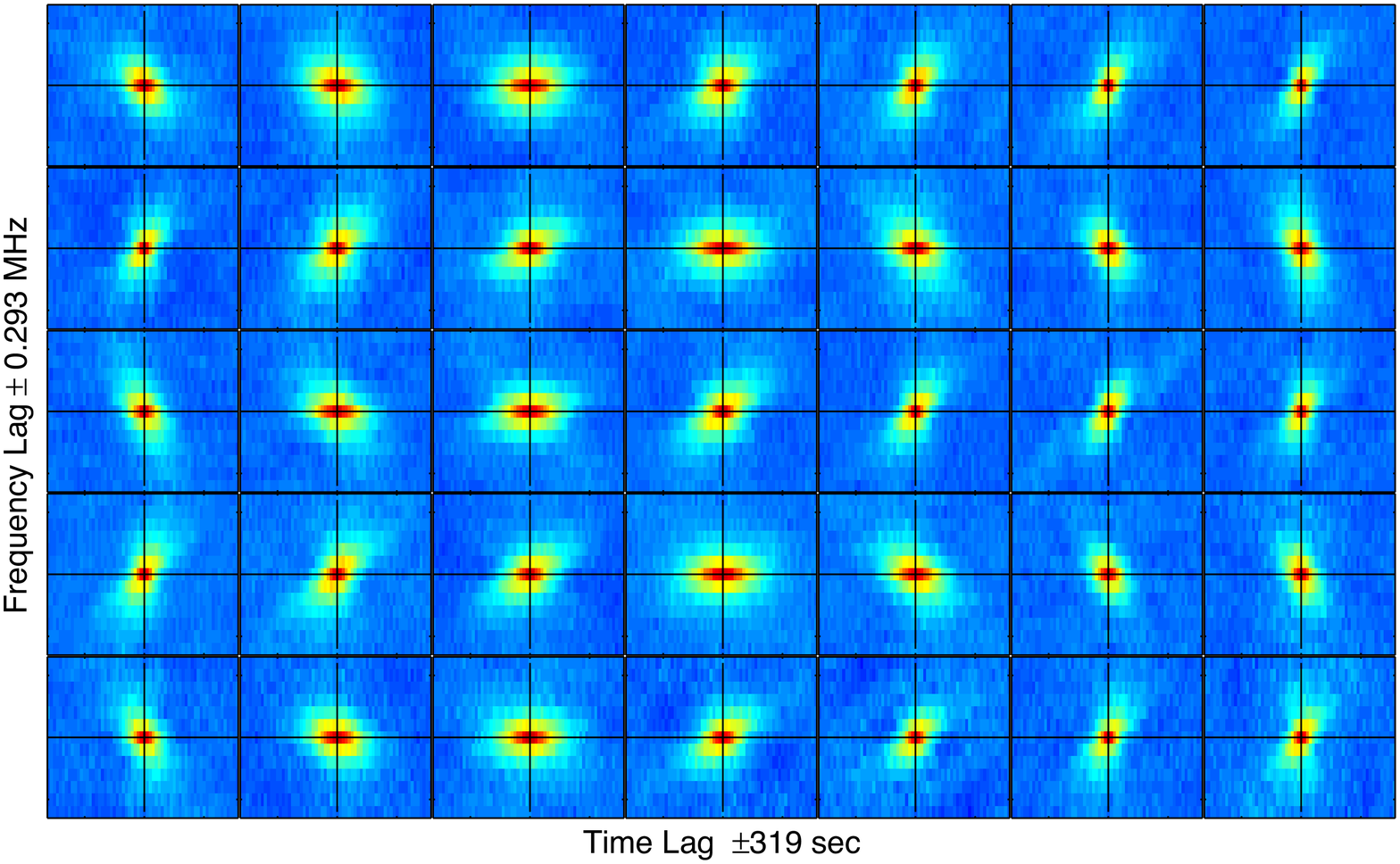}  &  \includegraphics[height=4.5cm, trim=100 0 700 0, clip]{colorbar.eps}\\
\includegraphics[ height = 4.5cm, clip]{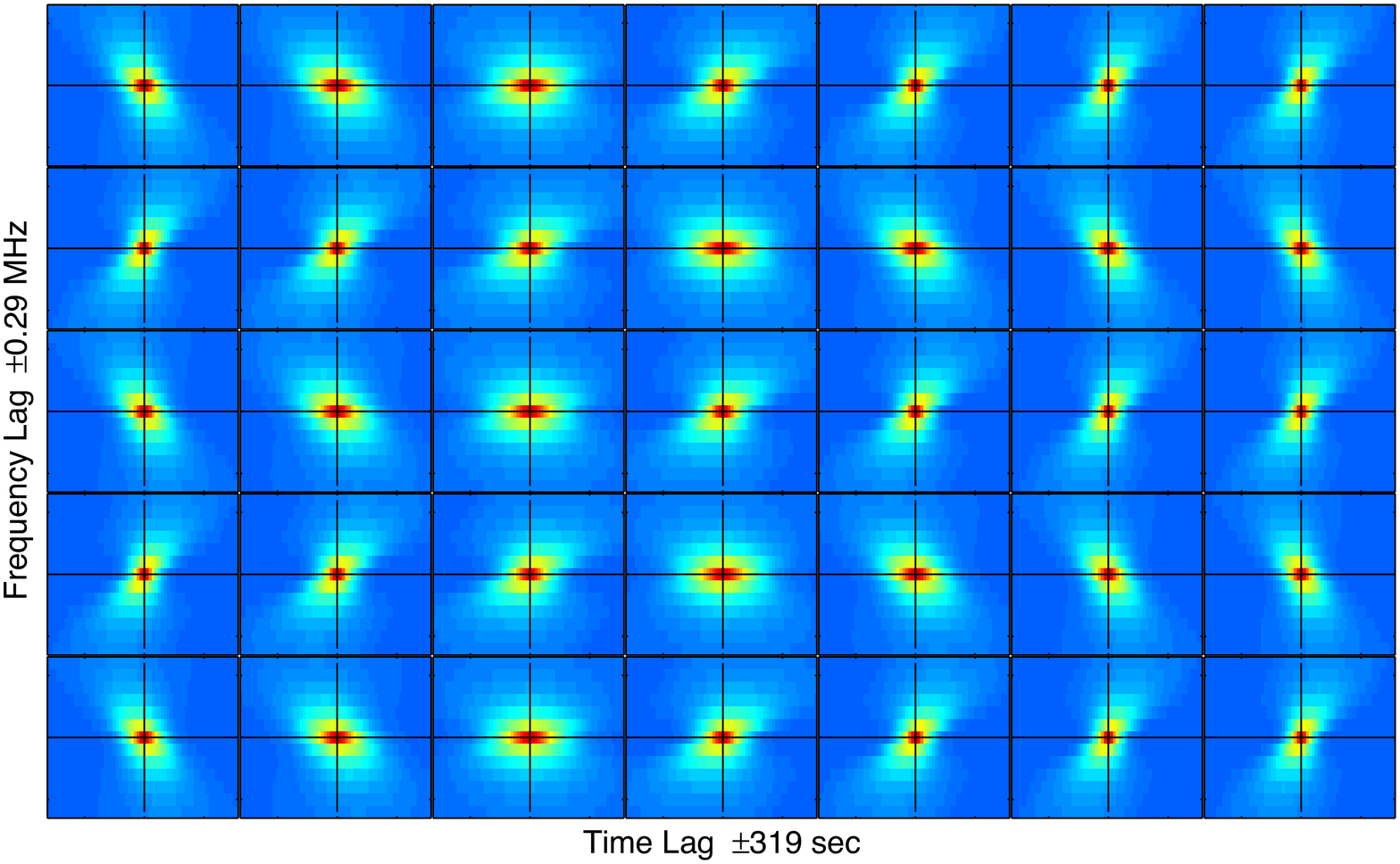}  &  \includegraphics[height=4.5cm, trim=100 0 700 0, clip]{colorbar.eps}\\
\end{tabular}
\caption{Frequency-time acfs for the 820 MHz dynamic spectra in the lower panel of figure \ref{fig:53560}.  
The time lag axis shows $\pm 319$ s with triangular tapering over the 320 s block length.
The format is the same as Figure 2 but there are 2.5 orbital cycles.}
\label{fig:ftacf_53467}
\end{figure}

The 820 MHz acf observations on the fifth day
are shown in the upper panel of figure \ref{fig:ftacf_53467} in the format
of figure \ref{fig:ftacf} with the fitted model in the lower panel.   
The data on all 5 days at 820 MHz were RFI free and gave good fits 
to the acf model, as we describe below.  

We precomputed a grid of models with anisotropies ($R=$ 0.05 0.22 0.38 0.6 0.8 0.88),
and at each $R$ we fitted the remaining 5 parameters with uniform weighting in the 
frequency-time plane. 
From the sums-of-squares of the residuals versus $R$ we fitted a parabola
to estimate the best-fitting value and its error. 
In Table \ref{tab:ftacfitL} we list these with the other fitted parameters
and their errors determined at the nearest grid value of $R$.
We also give an estimate of the reduced $\chi^2_{\rm red}$ (row 8), which requires  
the typical error at each point in the 2-d acf, which was calculated as follows. There are 
contributions from both system noise ($V_{\rm noise}/\sqrt{N_{\rm noise}}$) 
and from statistical variations in the ISS ($V_{\rm iss}/\sqrt{N_{\rm iss}}$) which sum
in quadrature.  While the noise term is independent in each of the 
pixels of the acf, the ISS term is correlated over the characteristic scales in frequency 
and time.  In the 820 MHz data the ratio of the variance in the ISS to the variance in the noise
was in the range $0.1-0.2$, such that the ISS and noise made roughly equal contributions to 
the acf error, which was typically $\sim 0.03$.  The resulting  $\chi^2_{\rm red}$
are near unity indicating that the fits are satisfactory.

The 5 anisotropies are somewhat lower than 
those estimated from the annual variation of the harmonics in Table 2, 
but are consistent at the 1.5$\sigma$ level.  However there is 
indication of temporal changes.   
The $\Delta\nu$ estimates are smaller than or comparable to the 49-kHz channel
bandwidth of the spectrometer, so they should be considered upper bounds.  
The $\siss$ estimates come directly from the model fit.
The interstellar velocities are derived from $u_x, u_y$ in the model fit via equation (\ref{eq:fg}),
where the listed velocity errors do not include effects of uncertainty in $s$ and $\Omega$.
As discussed in \S\ref{sec:annual} the proper motion velocity of the pulsar system that we used
could be too large.  However, even the lower estimate of  $\vpa=-5.3$, $\vpd=6.2$ km s$^{-1}$
would change the interstellar velocity by their constant vector difference (times $1-s$),
which is only about 4 km s$^{-1}$ in magnitude.
The last two quantities in Table \ref{tab:ftacfitL} are calculated from those above.  
The acf model at 1900 MHz in the lower panel of figure \ref{fig:ftacf} was fitted in a similar fashion, 
except that the axial ratio was held constant at $\ar=2$ ($R=0.6$) and the other
5 parameters were fit. The other observations at 1900 MHz also show ISS slopes and similar behavior.   
However the data are occasionally corrupted by RFI, 
which made the fitting unreliable.

\begin{table}[!ht]   
\caption{6 parameters (rows 2-7) were estimated for the 5 observing epochs at 820 MHz
from a grid search in $R$ with the fitting errors given assuming $R$ to be correct.
Errors in $V_{ISx},V_{ISy}$ do not include the small systematic errors due to uncertainty in $s$ and 
in the proper motion of the pulsar binary system; Rows 11 \& 12 are derived quantities.
}
\label{tab:ftacfitL}
\smallskip
\begin{tabular}{rrrrrr}
\tableline
Day  &	-3 &	211 &	311 &	379 &   467\\
\tableline
$R$ &.43$\pm$.02&.40$\pm$.04&.17$\pm$.05&.49$\pm$.02&.19$\pm$.03\\
$\psi_{AR}\;\deg$&48$\pm$3&35$\pm$3&32$\pm$7&53$\pm$4&47$\pm$5\\
$\sigma_{p} (10^9)$m &6.5$\pm$.7&11$\pm$1&9$\pm$1&13 $\pm$2&15$\pm$2\\
$\psi_{p}\;\deg$&-156$\pm$4&16$\pm$4&1$\pm$6&26$\pm$4&-6$\pm$1\\
$\Delta\nu_{iss}$(kHz)&47$\pm$2&50$\pm$2&44$\pm$2&37$\pm$2&63$\pm$2\\
$s_0 (10^6)$m&3.7$\pm$.4&3.5$\pm$.3&3.4$\pm$.3&3.9$\pm$.4&4.2$\pm$.4\\
$\chi^2_{\rm red}$&  1.6&  1.3&  1.3&  1.5&  1.3\\
$V_{ISx}$(km/s)& -1$\pm$1&-14$\pm$1&-12$\pm$1&-12$\pm$1&-21$\pm$1\\
$V_{ISy}$(km/s)& 32$\pm$1& 39$\pm$1& 41$\pm$2& 33$\pm$2& 29$\pm$2\\
$s_{ref}(10^{10}$m)&6.5$\pm$.6&5.7$\pm$.6&6.4$\pm$.6&8.7$\pm$.9&5.5$\pm$.6\\
$\theta_p/\theta_d$& 0.10& 0.19& 0.14& 0.15& 0.27\\
\tableline
\end{tabular}
\end{table}

\subsection{Reanalysis of Annual Modulation by the Earth's Motion}
\label{sec:reanal}

The acf analysis indicates that $\bdma{V_{IS}}$ and the anisotropy is somewhat time variable. 
Accordingly we modified our annual fitting routine for the harmonics to take variable $\bdma{V_{IS}}$ and anisotropy.
The model interpolates linearly between the values found above from the five 820 MHz acf epochs, to obtain
the appropriate values for 1900 MHz harmonics.  We included the formerly
discrepant data from MJD 52997 in the fits.
 
With variable anisotropy the reduced $\chi^2$ (3.6) was significantly worse than for fits with constant anisotropy. 
This probably reflects the fact that the mean $R$ from the 820 MHz acfs is somewhat lower than the
best fit value of the harmonic fit with constant $R$.  Thus evidence for variable anisotropy 
on a time scale of a year shown in Table \ref{tab:ftacfitL} is only marginal.

With variable $\bdma{V_{IS}}$ the fits showed a large improvement.
The fits with variable $\bdma{V_{IS}}$ are shown as magenta lines in
Figure \ref{fig:ksk0}. The revised model now fits the first and last points much better and is equally good in the
other points. We conclude that the evidence for variation in $\bdma{V_{IS}}$ on this time scale is
strong, and is the cause of the apparent discrepancy on MJD 52997.

We have redone the four harmonic fitting including MJD 52997 using variable $\bdma{V_{IS}}$
to produce ``best available'' values. In each case
we fit $s$, $\Omega$, $R$ and $\psi_{AR}$. We used the variable velocity model, or fixed 
velocities from the previous constant velocity fit. In two cases we held $\cos i = \pm0.0227$
and in the third case we fit $\cos i$, which yielded $i=88.1\pm 0.5\deg$.  
The results are given in Table \ref{tab:inc}. The case for $\cos i > 0$ is compelling as the alternative
location doubles the $\chi^2$.   It is remarkable how well the location of the
scattering medium and the alignment of the pulsar orbital plane are determined. This is because 
these parameters are not correlated with the inclination.

However, in the fit for $i$ the estimated anisotropy $R$ is significantly larger than the mean from
Table \ref{tab:ftacfitL}, and as noted, the fit is worsened by allowing it to vary.
This suggests that the uncertainty on the anisotropy is underestimated and weakens
the indication of temporal variation in the anisotropy.  However 
the velocity variation must be real and the variations, which are of the order
of $\pm10$ km s$^{-1}$, are probably not super-Alfv\'{e}nic. However, they suggest that the scattering
medium is much less homogeneous than has been assumed.

The fit for the inclination $i=88.1\pm 0.5\deg$ from the variable velocity model
agrees at the $1\sigma$ level with that assuming a fixed velocity in \S\ref{sec:inclin}.   
It  is also consistent with the confidence limits derived from the Shapiro delay at the 
1 $\sigma$ level for $i<90\deg$.    However the inconsistency in estimating $R, \psi_{AR}$,
discussed above, indicates a weakness in the model, which may contribute a systematic error 
that is not included in our estimate of $i$ and its standard error.

\begin{table}[htb]
\caption{Parameters estimated from fitting to
observations of the four normalized harmonic coefficients at 0.8 \& 1.9 GHz versus date
including MJD 52997 and using a variable IISM velocity model derived from the acf fits.
}
\label{tab:6params}
\smallskip
\begin{center}
\begin{tabular}{rrrr}
\tableline
\noalign{\smallskip}
Parameter & Fit i  & $i=91.3\deg$  &  $i=88.7\deg$   \\
\tableline   
$\cos i$ & 0.033 $\pm0.009$ & -0.0227 & 0.0227 \\
$s$ & 0.73 $\pm0.01$  & 0.74 $\pm.02$ & 0.72 $\pm0.01$ \\
$\Omega$ (deg) & $62\pm2$ & $63\pm3$ & $62\pm2$ \\
$R$  & $0.58\pm.08$  & $0.78\pm.12$ & $0.54\pm0.06$ \\
$\psi_{AR}$ (deg) &  $68\pm3$  & $75\pm3$ & $66\pm3$ \\
$V_{ISx}$ (km s$^{-1}$) & variable  & variable &  variable \\
$V_{ISy}$ (km s$^{-1}$) &  variable & variable & variable \\
$N_{\rm dof}$ & 63  & 64 & 64 \\
Reduced $\chi^2$& 2.6 & 5.0 & 2.7 \\
\tableline
\end{tabular}
\label{tab:inc}
\end{center}
\end{table}

\subsection{Anisotropy and Phase Gradient Variations}

\begin{figure*}[htb]
\hspace{5mm}
\includegraphics[trim= 120 100 70 100, clip, width=17cm]{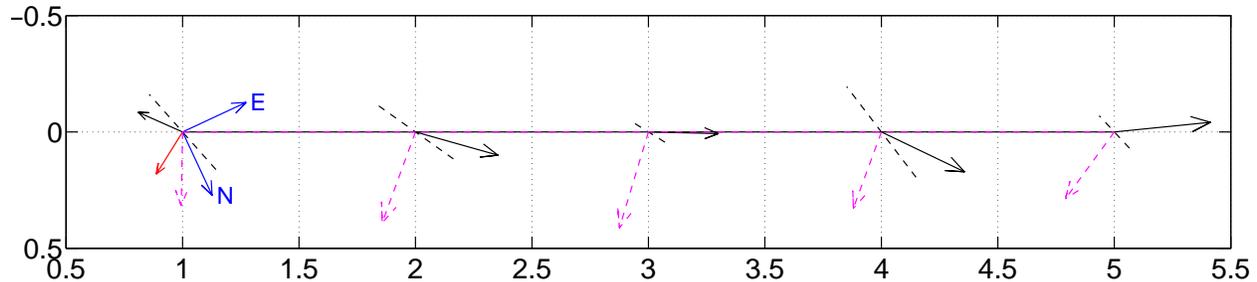}      
\caption{Vector representations of the IISM anisotropy, phase gradient and 
velocity in the plane of the sky for the 5 epochs observed at 820 MHz. 
The orbital $x$-axis (toward the line of nodes) is horizontally to the right
and the $y$-axis is vertically down. Note that with the orbit inclination
near $90 \deg$ its angular momentum vector points upwards in the figure.
Celestial North and East are marked by blue arrows on day 1.
The major axis of scattering is
shown as a black dashed line of length $R$ at angle $\psi_{AR}$ to the $x$-axis.  
The black solid arrow is the refractive displacement vector 
$\sigma_{p},\psi_{p}$, (parallel to the transverse phase gradient)
as defined in the Appendix (and arbitrarily scaled).   
The magenta dashed arrows show the interstellar velocity vector.
The system proper motion velocity is shown by the red arrow on day 1.  
The scaling of all velocity vectors is in 100 km s$^{-1}$ units. 
The Earth's motion round the Sun makes a cycloidal trajectory 
through the IISM, which complicates a spatial mapping of the 5 epochs.
} 
\label{fig:thetas}
\end{figure*}

The phase gradient for each of the five 820 MHz observations is plotted a solid line vector in
figure \ref{fig:thetas}. The anisotropy is plotted  as dashed line along its major axis on the
same figure. One can see that there is some variation in both
quantities.   Furthermore the mean direction of the phase gradient is nearly parallel
or anti-parallel to the mean major axis of the anisotropy.  Both quantities would vary randomly even if the IISM
were a uniformly turbulent Kolmogorov plasma from refractive effects due to the finite number of
scintles in the scattering disc.  Hence the variations in both quantities
in figure \ref{fig:thetas} could be due to such statistical variations. However the phase gradient does
show a persistent mean, which is not expected if the turbulence was statistically uniform.

\subsection{Theoretical model for phase gradient variations}
\label{sec:phgrad}

Since the phase gradient $\bdma{\nabla}\phi_{\rm p}$ 
for the 5 epochs at 820 MHz is well estimated, one can ask if the
gradients are typical of what one would expect of a Kolmogorov random
process, or if one must invoke a deterministic structure. This is easily
done for an isotropic medium and the anisotropy we have measured is
not large enough to make a significant difference.
The rms phase difference $\phi_{rms}(x)$ over a distance $x$ can be obtained from the
structure function of phase directly as $\phi_{rms}(x) = D_{\phi}(x)^{0.5}$ and
the rms gradient would be $\phi_{rms}(x)/x$. Here we
have measured $\bdma{\nabla}\phi_{\rm p}$ over $s_{ref}$ so we can compare
it with $D_{\phi}(s_{ref})^{0.5} / s_{ref}$. However it is more intuitive to
compare the resulting angular displacement $\bdma{\theta_p}$ with the rms
scattering angle $\theta_0 = 1 / k s_0$ where $k = 2 \pi/\lambda$ is the
propagation constant. Substitution yields
\be
\mbox{rms}(\bdma{\theta_p} ) / \theta_0 = (s_0 / s_{ref})^{1/6}.
\ee
The scales $s_0$ and $s_{\rm ref}$ are listed in 
Table \ref{tab:ftacfitL} where the bottom row shows that 
$\theta_p/\theta_{0} \sim 0.15\pm0.05$.  With $s_0/s_{ref} \sim 5\times 10^{-5}$, 
the predicted rms value is 0.2 on a refractive timescale 
$10-20$ days.   Thus with observations separated by 60-100 days, 
we are seeing independent samples of a phase gradient which is close
to the rms expected from a Kolmogorov spectrum. 
However it should be noted that comparable phase gradients, lasting for decades, have
been observed in other pulsars \citep{keith}. These are due to transverse gradients
in the electron column density (dispersion measure), which may be due
to a breakdown in the homogeneity of the turbulence or to
the presence of a discrete plasma structure somewhere along the line of sight.

\subsection{Theoretical model for anisotropy variations}
\label{sec:anisot}

The scattering from an isotropic Kolmogorov phase screen will, in any particular
realization, appear slightly anisotropic simply because there are a finite number
of ``scintles'' in the scattering disc. This effect has been discussed and quantified
by \citet{rnb}.
From their Table 1 one can find an rms 
value $R \sim 0.7 (s_0/s_{ref})^{-1/6}$. If the IISM were isotropic, then from our
Table \ref{tab:ftacfitL} (lines 6 \& 7) we would expect an rms $R \sim 0.14$. We
actually observe $0.2 < R <  0.5$, which suggests that $R = 0.3 \pm0.14$ would
be a realistic estimate. In this case the variation of $\pm0.14$ is simply statistical
variation in a Kolmogorov medium with a constant anisotropy.

\section{The cross correlation between the ISS of the two pulsars}
\label{sec:rhoab}

In Paper 1 we measured the correlation between the ISS of the two
pulsars near the time of A's eclipse, which led to an estimate
of the orbital inclination that was even closer to edge-on than
found from the timing observations \cite{kramer06}.  
However, as Perera et al (2010) have shown, the time windows
during each orbit where B is observable have shifted and shrunk
due to relativistic precession of B's spin about the orbital angular 
momentum vector. By 2008 the B pulsar had become undetectable.

We were only able to measure the cross-correlation 
in the ISS of the pulsars for 6 of the observing epochs. 
In our interpretation we assume that, where the projected paths of
the two pulsars cross, their lines of sight through the IISM
to the Earth are identical.  In such a situation ISS would cause
an identical modulation of intensity versus frequency.
Because of the velocities of the center of mass and the Earth 
relative to the IISM,  the paths cross at different times for the two pulsars,
but the ISS will be highly correlated if the IISM changes slowly enough.

\begin{figure}[tb]
\begin{tabular}{ll}
\includegraphics[trim = 15 25 0 35, height=4cm, clip]{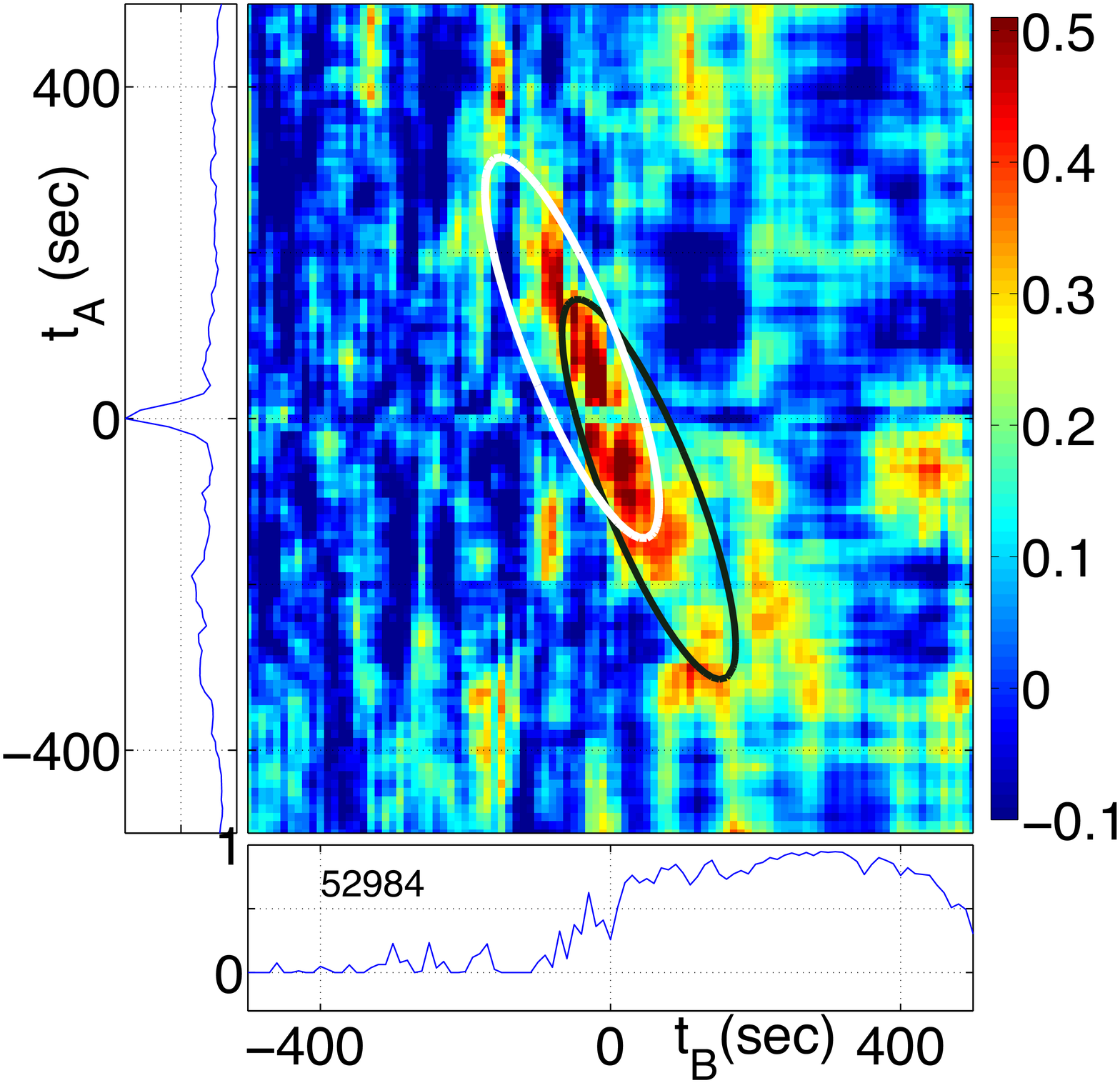} & 
\includegraphics[trim = 15 25 0 35, height = 4cm, clip]{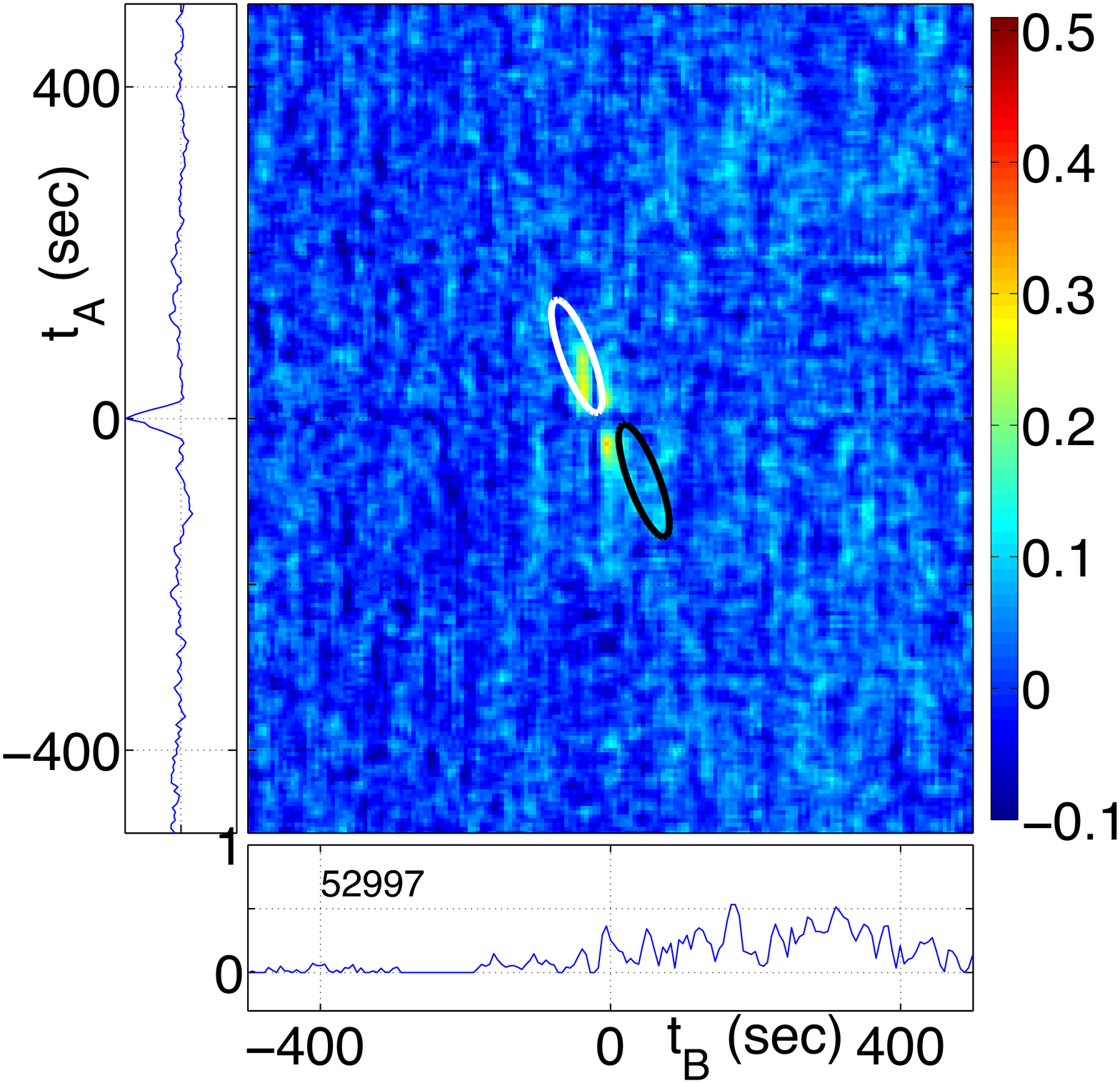} \\
\includegraphics[trim = 15 25 0 35, height=4cm, clip]{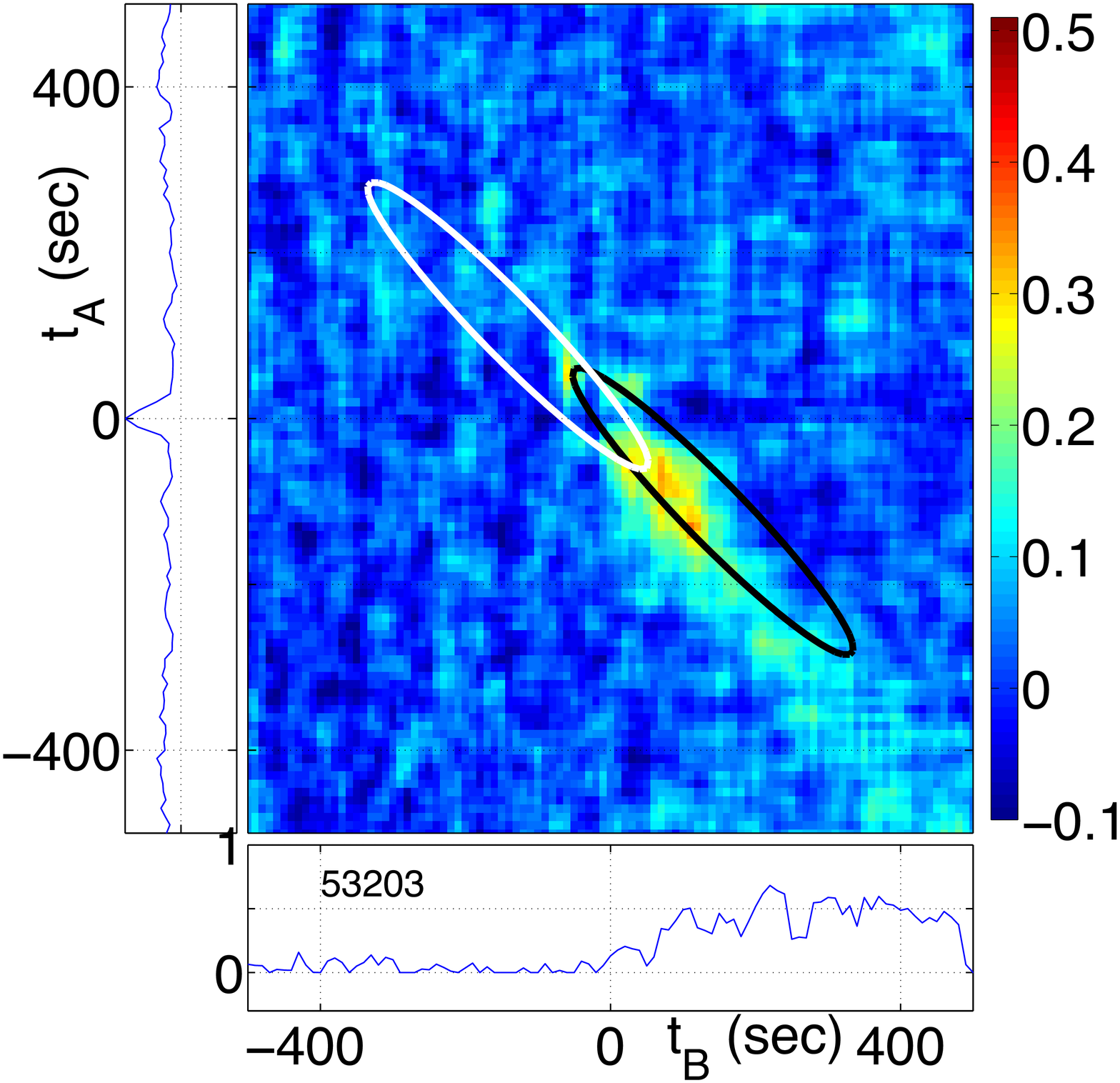} & 
\includegraphics[trim = 15 25 0 35, height = 4cm, clip]{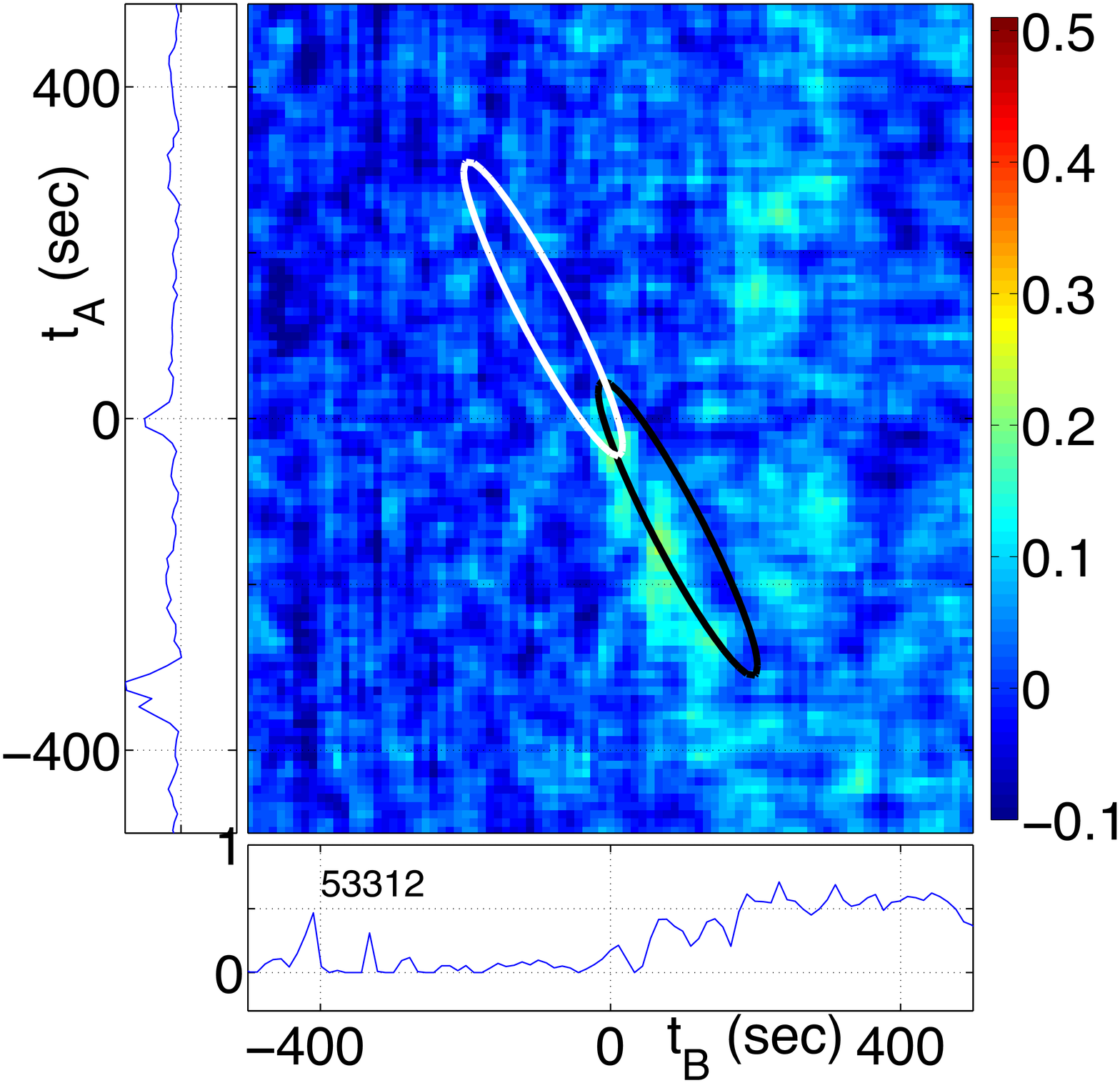} \\
\includegraphics[trim = 15 25 0 35, height=4cm, clip]{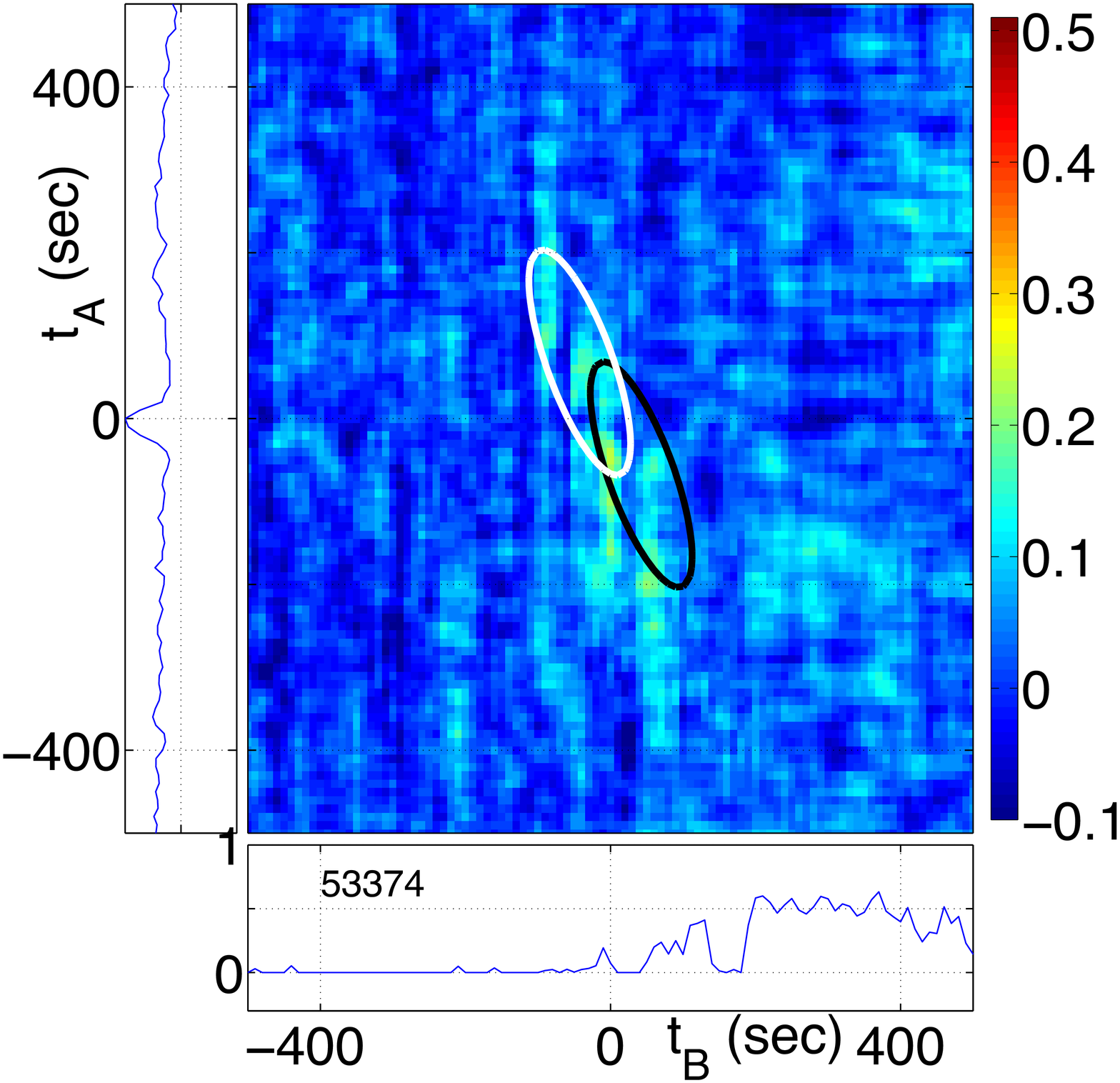} & 
\includegraphics[trim = 15 25 0 35, height = 4cm, clip]{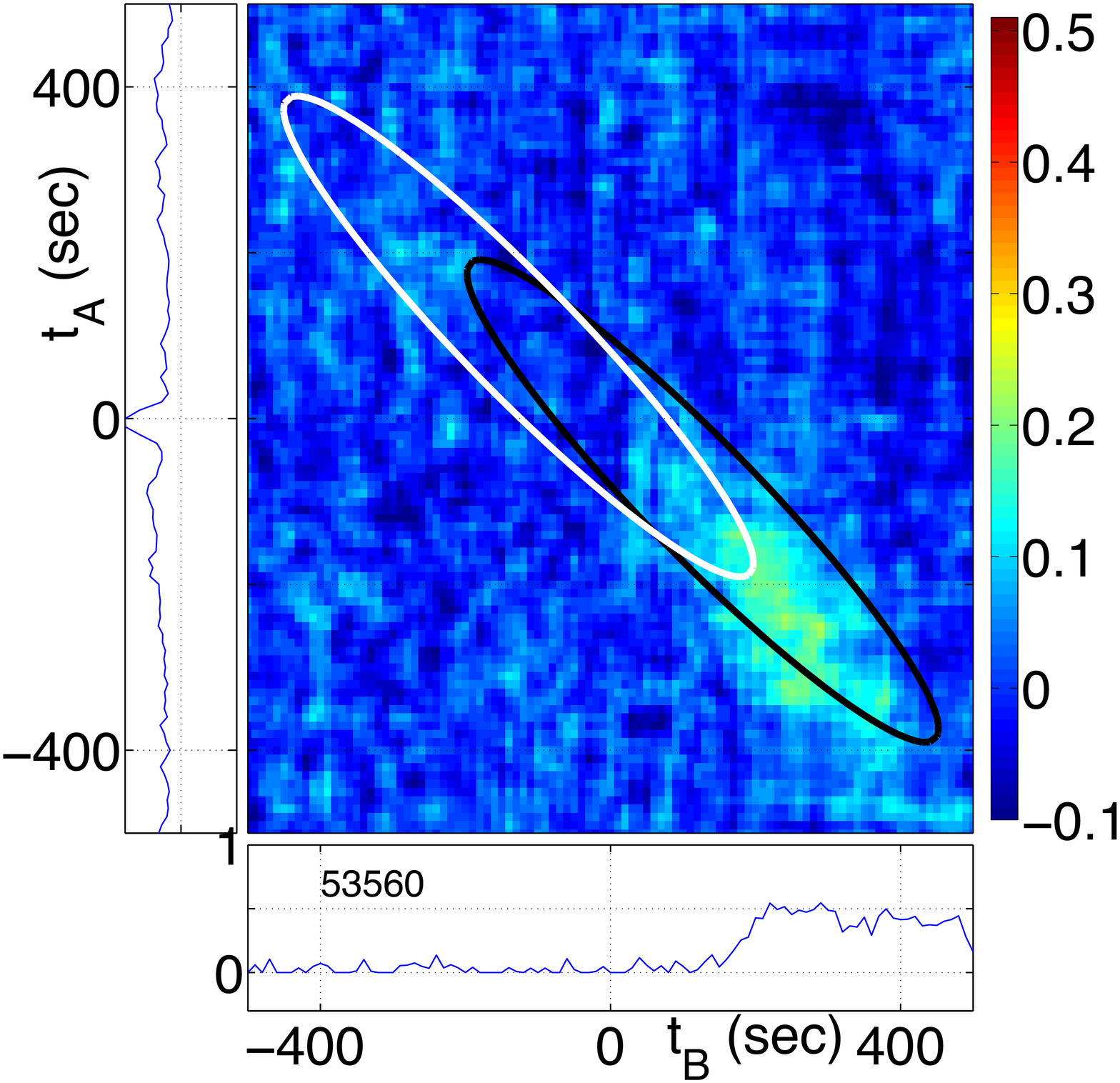} 
\end{tabular}
\caption{Correlations $\rho(\ta ,\tb)$ between the ISS of pulsars A and B
on MJDs 52984, 52997, 53203, 53312, 53374, and 53560
at frequencies of 1420, 820, 1950, 1950, 1950, and 1950 MHz working left to right
and top to bottom.   The plotted coordinates
are observing times ($\ta,\tb$) relative to the time of A's eclipse.
The 50\% contour of the theoretical models described in the text are
shown as white and black ellipses for $i=89.0\deg$ and $i=91.0\deg$, respectively. 
The models will be reduced by the factors, shown in the marginal plots,  due to the
changing signal-to-noise ratios for A and B. The scale in the marginal plots is 0 to 1.0.}
\label{fig:rhoab}
\end{figure}

As in paper 1 we computed the temporal correlation $\rho(\ta ,\tb)$
by subtracting the mean and cross-correlating the intensity over frequency
from A and B at time offsets $\ta$ and $\tb$, relative to the center
of A's eclipse.  The results are shown in Figure \ref{fig:rhoab} 
for the 6 days that showed significant correlation.
$\rho(\ta ,\tb)$ is a measure of the correlation between the
ISS of the two pulsars $\rho({\bf s} = \ra -\rb)$
where $\ra = \va \ta$ and $\rb = \vb \tb + (0, \ybz)$, with the 
velocities projected transverse to the path through the IISM.
Here A's position at its eclipse defines the origin of the $x,y$ coordinates.
$(0, \ybz)$ is the projected offset of B, where $\ybz = -d_{AB}\cos i$ 
with $d_{AB}$ their separation at the eclipse.
The time offsets ($\ta$ and $\tb$)
are short enough that $\va$ \& $\vb$ are effectively constant.
The observations show greatest correlation when A
is before the eclipse and when B is after the eclipse.

We have overplotted two 50\% correlation contours derived from the model developed in previous sections.
The two models correspond to orbital inclinations of $89.0\deg$ and $91.0\deg$ and are plotted as
overlapping white and black lines, respectively.
Since the observed correlations are normalized by the square root of the product
of the total variances in the A and B spectra, the model correlations must be reduced by factors that depend 
on the signal to noise ratios.  The marginal plots of these factors
show that the A pulsar flux is steady except
when it is eclipsed. However B only turns on for positive $\tb$,
and so correlation is only observed when B is near or after the eclipse.
Thus although we were able to detect  the correlation on 6 days, there is only one
day on which the peak correlation can be measured with any confidence
in Figure \ref{fig:rhoab}.   Only in the first (upper left) panel does
B have any significant flux before the eclipse and it is consistent 
with the white theoretical contour.  
In the other panels there is no flux in B inside most of the predicted 50\% white contour. 
However where there is measurable B flux the correlation level for the
white ellipse is consistent with the prediction. 
Conversely the black ellipses, ($i=91\deg$), are shifted well into the 
$t_B$ region, yet the correlation is much
lower than predicted by the model.
In the other observing epochs we observed observed no flux from B
for $\tb<0$ and no AB correlation. Had the inclination been $> 90\deg$
we should have seen AB correlation.

We have not been able to make a satisfactory estimate for the bias
to the correlation due to B's absence before the eclipse and so we
cannot define a formal error on the inclination from this investigation.
It is clear however that the AB correlations favour the higher end of the range
 $i=88.1\pm0.5\deg$ found in \S\ref{sec:inclin}.  As the inclination
 is decreased from $89\deg$ the white contour in Figure \ref{fig:rhoab}
moves to the upper left further from the measured positive correlation.
Thus we conclude that the measured temporal correlation 
$\rho(\ta ,\tb)$ is consistent with the model developed earlier and supports
the conclusion that the inclination is  $\sim1\deg$ less than $90\deg$.   However,
it does not have the power to improve the accuracy of that model.

\section{Conclusions}
\label{sec:conclusions}

Although we were not able to achieve all the original objectives of this experiment, we have been
able to determine the orientation of the orbital plane of the system ($\Omega = 65\pm2\deg$ see Table \ref{tab:inc}) 
and the inclination of the orbit ($i=88.1\pm0.5\deg$). The inclination of the orbit is
consistent with earlier measurements of the Shapiro delay, if the smaller of the two possible
Shapiro delay solutions ($88.7\deg$) is taken. 

Knowing the orientation of the system relative to its proper motion will allow immediate progress on other scientific problems.  The first of these is the nature of the supernova that created the B pulsar.  Several attempts have been made to determine the magnitude of the kick to the nascent B neutron star (e.g., \citet{piran, willems,stairs06});
these involve tracing the path of the binary system back to possible birthplaces in the Galactic Plane, evolving the orbital eccentricity and semi-major axis to the appropriate age, and determining which set of kick magnitudes and directions is compatible with the known properties of the system. Most of these studies point to a small kick and therefore likely very little tilt of the post-supernova orbital plane relative to the pre-supernova one; this is reinforced by the finding that the spin axis of the A pulsar is close to aligned with the orbital angular momentum \citep{ferdman}.  Previous work on the evolution of this system did not use a constraint on $\Omega$, as the anisotropy of the IISM prevented a robust determination of the angle at the time \citep{stairs06}
Recent modeling \citep{wong} has found that kicks in the plane of the pre-supernova orbit are preferred over polar kicks; the rather large angle (~60$^{\circ}$) found here between the Line of Nodes and the proper motion may be at odds with this result.  Revised modeling incorporating the constraint on $\Omega$ will be presented elsewhere.

The determination of $\Omega$, plus the resolution of the sign ambiguity in $\cos i$, permits a strengthening of the double-pulsar test of preferred-frame effects in semi-conservative theories of gravity \citep{wex}. Such effects would produce periodic changes in the longitude of periastron and the eccentricity of the system, in a manner that depends on the orientation of the system relative to the coordinates of the preferred frame. Therefore, the knowledge of the orientation will allow a more precise limit to be set on the parameters of this theory.

We model the dominant interstellar scattering plasma as a thin layer located at a distance from the pulsar of 
$73\pm1$\% of the distance to the Earth.  The success of this ``thin screen'' model emphasizes 
the highly localized distribution of scattering plasma along the line of sight.
We also measured its velocity and scattering parameters. 
The velocity is about 40 km s$^{-1}$ with respect to the Sun,
with variations of about 10 km s$^{-1}$, that are  comparable with expected Alfv\'{e}n speeds.
The level of turbulence varied by a factor of two on a time scale of months, much greater than the statistical variation
expected due to refractive effects in a homogeneous Kolmogorov random process.
At 5 epochs we measured its anisotropy (axial ratio $1.2 - 1.7$)
and phase gradient (due to a transverse gradient 
in the electron column density).
There is some significant variability between epochs in both of these parameters;
however the variations are at the level one might expect in 
different refractive realizations of a homogeneous Kolmogorov random process
concentrated in a thin layer.  

Our results add to the evidence suggesting that the IISM model 
of homogeneous isotropic Kolmogorov turbulence is no longer adequate.
There is accumulating evidence for: anisotropy and intermittency in the turbulence
on sub AU scales \citep{ric02,dt03,tunstov2013,Hill,brisken} 
and for persistent phase gradients \citep{keith}. 
Evidently this default model of turbulence in the IISM will need to be modified. 
Apart from the light this throws on the interstellar plasma, turbulence in the IISM is a
problem for accurate pulsar timing which is limited in precision by dispersion and scattering \citep{keith}.

\acknowledgments
The National Radio Astronomy Observatory is facility of the National Science
Foundation operated under cooperative agreement by Associated
Universities, Inc.  Coles and Rickett acknowledge partial support
from the NSF under grant AST 0507713. Rickett thanks the Cavendish 
Astrophysics Group at Cambridge University for their hospitality. Pulsar
research at UBC is supported by an NSERC Discovery Grant.

\appendix

\section{Space-frequency correlations in the limit of strong diffractive scintillation}
\label{app:ftacf}

The theory for $C_I (\vsigma,\Delta\nu)$
was reviewed by \citet{LR99} and we follow their treatment here. 
The correlation of intensity is a fourth order moment of scattered
electric field.   While in general fourth moments cannot be solved
analytically, they simplify in the limit of strong diffractive scintillation
to the square of second order moments of electric field versus 
spatial offset $\vsigma$ and frequency difference $\Delta \nu$.  
Here we assume scattering in a thin layer of plasma, i.e.
a ``phase screen''.  As noted $C_I (\vsigma,\Delta\nu)$ is given by
\be
C_{I}(\vsigma,\Delta \nu) = | \Gamma_{D}(\vsigma,\Delta \nu)|^2  \; ,
\label{eq:CD}
\ee
where $\Gamma_{D}(\vsigma,\Delta \nu)$ is the diffractive part of
the electric fields covariance. Here $\vsigma$ is defined
as a transverse distance measured at the screen.
The definition of ``diffractive'' here involves factoring out
a term that corresponds to fluctuations of the pulse arrival time.
We assume spherical waves from a pulsar
at distance $z_p$ beyond a scattering screen which is
at a distance $z_o$ from the observer.  The pulsar distance is thus $L = z_o+z_p$.
Equation (59) of \cite{LR99} gives:
\begin{equation}
\Gamma_{D}(\bdma{\sigma},z_e;\nu_m,\Delta \nu)   = {\displaystyle
\frac{\nu_m^2}{2\pi i z_e c \Delta \nu}
\int \!\! \int_{-\infty}^{\infty}
\exp\left[ -0.5 D_{\phi}(\bdma{\sigma}';\nu_m) \right]} \;
{\displaystyle
\exp\left[ i\frac{\nu_m^2}{2 z_e c \Delta \nu}
            \left|\bdma{\sigma}-\bdma{\sigma}'\right|^2
    \right]
\, d^2 \bdma{\sigma}' 
\, \mbox{,}
}
\label{eq:GamD1}
\end{equation}
Here $z_e = z_p z_o/L$ and $D_{\phi}(\bdma{\sigma}';\nu_m)$
is the structure function of the plasma phase caused by the screen
at frequency $\nu_m$ and spatial separation $\bdma{\sigma}'$,
as given in equation \ref{eq:dphi}.

We change  the spatial coordinate to $\bdma{\sigma}'' = \bdma{\sigma}' -  \bdma{\sigma}$
and introduce normalized variables:
\be
p = [|\bdma{\sigma}''|/s_0]^2  \; \; \mbox{and} \;\;
v = ({\Delta \nu}/{\nu_m})({r_{{\rm Fe}}^2}/{s_0^2})   \;\;  \mbox{where} \; r_{{\rm Fe}} = \sqrt{z_e/k_m} \; .
\label{eq:sv}  
\ee
  The second moment can then be written in circular coordinates as
\be
\Gamma_{D}(\bdma{\sigma},z_e;\nu_m,\Delta \nu) &=& \frac{1}{i 4\pi v} \int_{0}^{\infty} e^{ip/2v} 
\int_{0}^{2\pi} exp[-0.5(\alpha+\beta\sqrt{p}+\gamma p)^{5/6}] d\theta'' \; dp ,
\label{eq:GamD2} \\
\mbox{where \;\;} \alpha &=& (a \sigma_x^2 + b \sigma_y^2 +c\sigma_x\sigma_y)/s_0^2  \nonumber  \\
\beta &=&  - (2a \sigma_x\cos\theta'' + 2b \sigma_y\sin\theta'' +c(\sigma_x\sin\theta''+\sigma_y\cos\theta''))/s_0 \\
\gamma &=& a \cos\theta''^2 + b \sin\theta''^2 +c\sin\theta''\cos\theta'' \nonumber 
\ee 
The $\theta''$ integral is simple to do numerically, 
but the Fourier-like integral over $p$ requires 
care as the normalized frequency offset $v$ approaches zero.  

Note that the acf versus frequency offset at
a single antenna is included as $\bdma{\sigma}=0, \alpha=0, \beta=0$.
So we used equation (\ref{eq:GamD2}) to compute the normalized 
frequency decorrelation width, $v_{0.5}$, versus axial ratio, holding $s_0$ constant.  The result
is that the higher the axial ratio the narrower the width $v_{0.5}$;
for example, an axial ratio 4:1 reduces $v_{0.5}$ to 0.43
compared to 0.96 for an axial ratio of 1:1.   Nevertheless,  the shape
of the acf is only weakly dependent on the axial ratio.  Consequently
in reporting the decorrelation bandwidths, we fitted the isotropic model and
record the frequency offset for a 50\% reduction in the acf,
since the bias by unknown axial ratio is less than the
fitting error.

\subsection{Effect of refraction and application to the ISS of pulsar A}
\label{app:ref}

A constant phase gradient over the scattering disc will cause a refractive shift of the entire
diffraction pattern by an angle $\bdma{\theta_p}$ as given in equation (\ref{eq:thetap}). This gives rise
to a displacement $\bdma{\sigma}_{\rm p} = z_e \bdma{\theta}_{\rm p} \propto \nu^{-2}$. The
frequency derivative $d\bdma{\sigma}_{\rm p}/d\nu = -2 \bdma{\sigma}_{\rm p} / \nu$. So for
a small refractive shift $\Delta \bdma{\sigma}_{\rm p}$ over a small frequency range $\Delta \nu$
we can write $\Delta \bdma{\sigma}_{\rm p} \sim  -2 \bdma{\sigma}_{\rm p}(\Delta \nu/\nu)$.
The time/frequency correlation $C_I (\tau,\delta\nu)$ can then be written
\be
C_I (\tau,\delta\nu) = C_I (\bdma{\sigma} =\bdma{V_{\rm A}}t\,(z_o/L) -2 \bdma{\sigma}_{\rm p} (\Delta \nu/\nu),\Delta\nu).
\ee
Models computed in this way are shown in the 28 sub-panels of the lower plot of figure 
\ref{fig:ftacf}.  Time lag is plotted horizontally and frequency lag vertically.
The slopes come from a constant refractive shift which appears to reverse in sign due
the changing pulsar velocity over its 2.45 hr orbit.   

\section{Influence of anisotropy on frequency-time acf}

\begin{figure}[htb]
\center{
\begin{tabular}{rl}
\includegraphics[width=6.5cm]{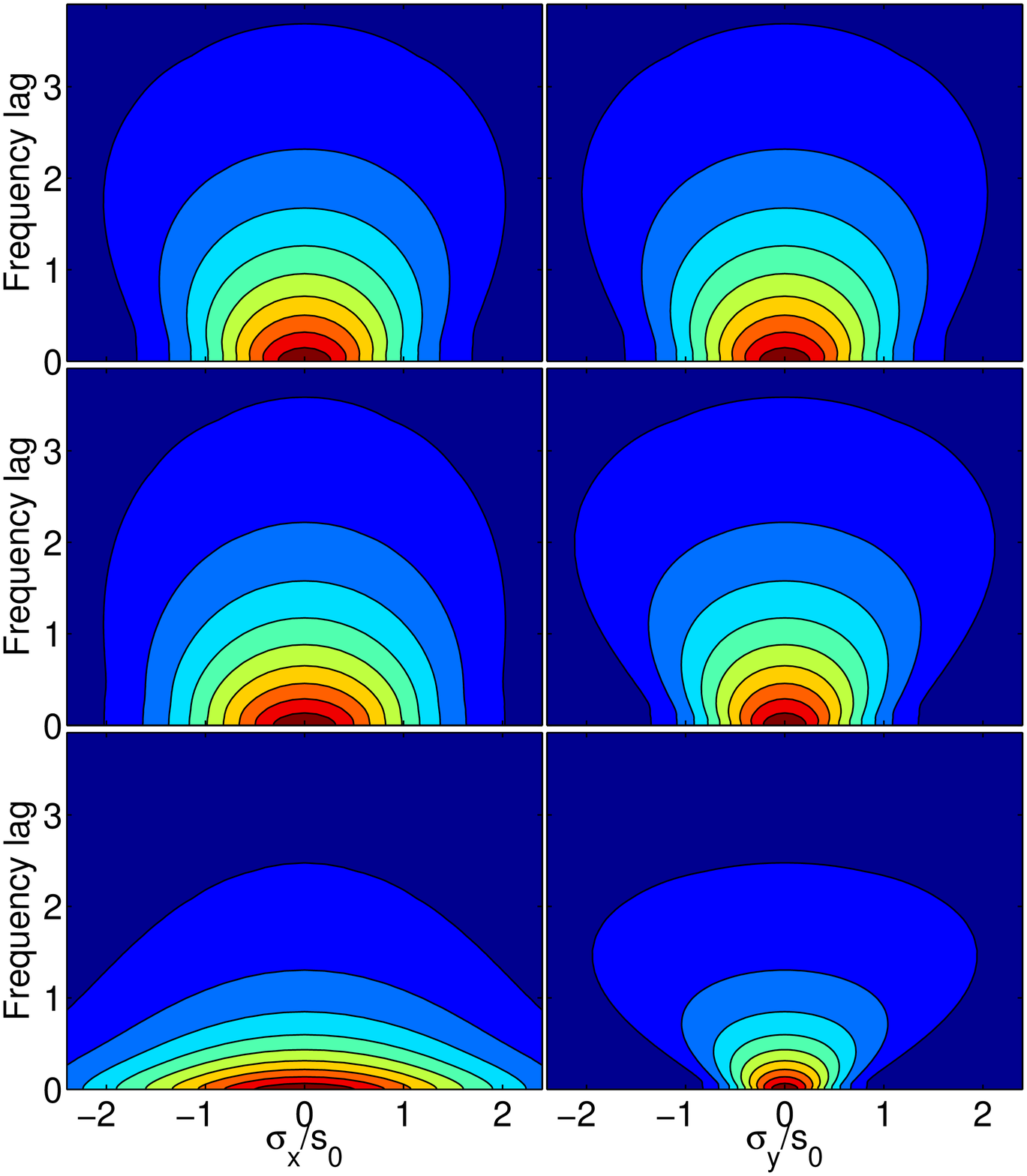} 
&   \includegraphics[width=6.5cm]{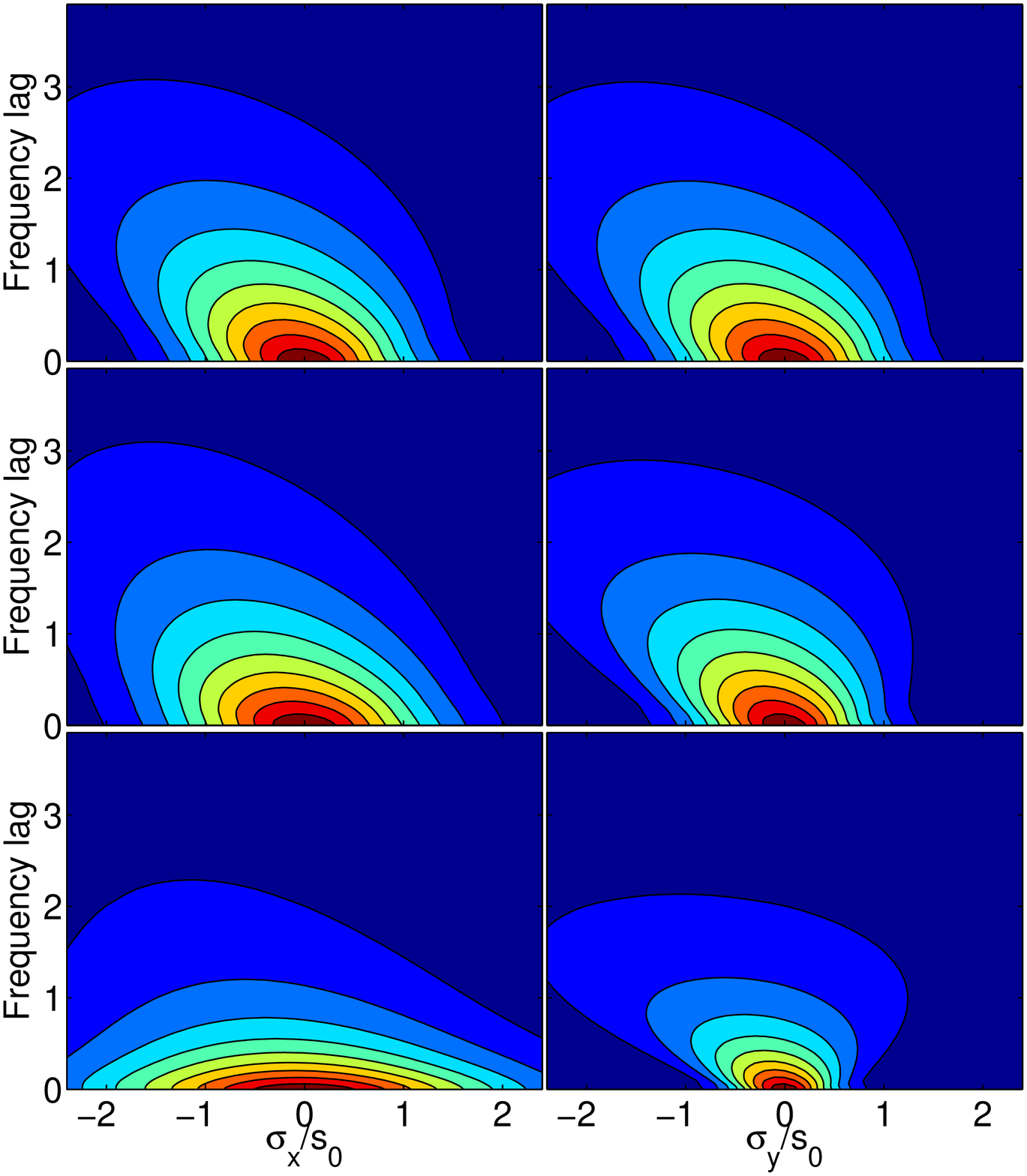} \\  
\end{tabular}}
\caption{ Theoretical acf $C_{D}(\vsigma,\Delta \nu)$ for three levels of anisotropy
($R=0.05, 0.38, 0.88$).  Spatial and frequency scales are normalized in units of the 
corresponding diffractive scales versus $\sigma_x,\sigma_y$. The major axis is along $\sigma_x$.
Contour levels are at 0.1 through 0.9 in steps of 0.1 and each acf is  symmetric about the origin.
\it Left: \rm no refraction. \it Right: \rm refractive phase gradient
$\theta_{px}=\theta_{py}=\theta_d/2$}.
\label{fig:fdc3_0}
\end{figure}

The earlier work of \citet{LR99} was done for thick and thin screens, but only for isotropic
scattering. Here we extend their work to a thin anisotropic screen with a refractive gradient
which is constant over the scattering disc. They showed that the
shape of contours of constant correlation changed significantly with the level of correlation
and this effect was more prominent in thin screens than in thick screens.  Here we find that
the effect is even more prominent as the anisotropy increases. While the higher level contours 
near the peak are approximately elliptical,  at lower levels they bulge outwards in the spatial 
coordinate as the frequency offset increases. Of course this is connected with the parabolic
arcs which can be seen in the fourier transform of $C_I (\tau,\Delta\nu)$ \citep{stinebring01}.

Cuts through $C_I (\vsigma,\Delta\nu)$ for $\vsigma = (\sigma_x,0)$ (left) and 
$\vsigma = (0, \sigma_y)$ (right) are shown in Figure \ref{fig:fdc3_0} in the case of
no phase gradient (left) and a phase gradient for which $\theta_{px}=\theta_{py}=\theta_d/2$ (right).
One can see that the distortion in the shape of the contours persists in the presence of a
refractive gradient, that it increases with increasing anisotropy (downwards in the Figure), and
that it is quite different in the $x$ and $y$ planes. It is these features that give the frequency-time acfs the potential to estimate both anisotropy and refractive gradient.

\subsection{Fitting the model acfs}

We fitted a model $C_I (\tau,\Delta\nu)$ to the set of observed acfs on a given 
date of observation.  For each data block, centered at orbital phase 
$\phi$, we start with the normalized velocities in equation (\ref{eq:uxuyw}) and 
write $\vsigma (\tau,\Delta\nu)$ as
\be
\sigma_x (\tau,\Delta\nu) = (z_o/L) V_o \tau \,(u_x-\sin\phi) - 2 \sigma_{px} \nu^{-1} \Delta\nu \;\; \mbox{and} \;\;
\sigma_y (\tau,\Delta\nu) = (z_o/L) V_o \tau \,u_y \sqrt{a/b} - 2 \sigma_{py} \nu^{-1} \Delta\nu  \; .  \nonumber
\ee
Here the x-component varies linearly with $\sin\phi$, but
the y-component is independent of orbital phase, because with $\cos i$ is so close to zero that
$V_{Ay}$ depends only on the center of mass velocity.   Thus the combination of
the orbital and center of mass motion of the pulsar gives a spatial offset vector
$(\sigma_x, \sigma_y)$ that swings over a range in angles governed
by parameters $u_x,u_y$. It is this that provides the sensitivity of the ftacf to 
spatial orientations.

Then we simply substitute the $\vsigma$ obtained
above in equation (B1). To speed execution we precompute the 3-dim $C_I (\vsigma,\Delta\nu)$ over
a 3-dim grid and find the necessary values of $C_I (\vsigma (\tau,\Delta\nu), \Delta\nu)$ by interpolation.
The 8 parameters to be determined are:
$\Delta\nu_{iss}$ that connects $\Delta \nu$ to $v$ in 
equation (\ref{eq:sv}); $u_x; u_y; R; \psi_{AR}; \sigma_p; \psi_p$ \& $V_o/s_0$.   
$u_x, u_y$ are partially  constrained by the fitting of harmonic coefficients 
to $\tiss(\phi)$.  We proceeded in an iterative fashion also constrained by 
the results of the annual fitting described in \S\ref{sec:earthmod}.

\end{document}